\documentclass[%
 reprint,
 superscriptaddress,
 amsmath,amssymb,
 aps, 
 pra,
floatfix,
]{revtex4-2}

\usepackage{graphicx}% Include figure files
\usepackage{dcolumn}% Align table columns on decimal point
\usepackage{bm}% bold math
\usepackage{comment}
\usepackage{color} 
\usepackage{url}

\usepackage{braket}
\usepackage{booktabs}
\usepackage{threeparttable}
\usepackage{float}
\begin{document}

\preprint{APS/123-QED}

\title{\textbf{Analytical and numerical studies of periodic superradiance} 
}% 

\author{Hideaki Hara}
 \email{hhara@okayama-u.ac.jp}
 \affiliation{Research Institute for Interdisciplinary Science, Okayama University, Okayama, 700-8530, Japan}
\author{Yuki Miyamoto}
 \email{miyamo-y@cc.okayama-u.ac.jp}
 \affiliation{Research Institute for Interdisciplinary Science, Okayama University, Okayama, 700-8530, Japan}
\author{Junseok Han}
 \affiliation{Research Institute for Interdisciplinary Science, Okayama University, Okayama, 700-8530, Japan}%
 \affiliation{Department of Physics and Astronomy, Seoul National University, Seoul, Korea}%
\author{Riku Omoto}
 \affiliation{Research Institute for Interdisciplinary Science, Okayama University, Okayama, 700-8530, Japan}%
\author{Yasutaka Imai}
 \affiliation{Research Institute for Interdisciplinary Science, Okayama University, Okayama, 700-8530, Japan}%
\author{Akihiro Yoshimi}
 \affiliation{Research Institute for Interdisciplinary Science, Okayama University, Okayama, 700-8530, Japan}%
\author{Koji Yoshimura}
 \affiliation{Research Institute for Interdisciplinary Science, Okayama University, Okayama, 700-8530, Japan}%
\author{Motohiko Yoshimura}
 \affiliation{Research Institute for Interdisciplinary Science, Okayama University, Okayama, 700-8530, Japan}%
\author{Noboru Sasao}
 \email{sasao@okayama-u.ac.jp}
 \affiliation{Research Institute for Interdisciplinary Science, Okayama University, Okayama, 700-8530, Japan}%

\date{\today}% It is always \today, today,

\begin{abstract}
We conduct a theoretical study to understand 
the periodic superradiance observed in an Er:YSO crystal.
First, we construct a model based on the Maxwell-Bloch equations 
for a reduced level system, 
a pair of superradiance states and a population reservoir state.
Analysis of the eigenvalues of the linearized differential equations shows that periodic superradiance can be realized only for certain parameters.
We also derive two-variable equations 
consisting of the coherence and population difference 
between the two superradiance states, 
which contain the essential feature of the periodic superradiance.
The two-variable equations 
clarify a mathematical structure of this periodic phenomenon 
and give analytical forms of the period, pulse duration, 
and number of emitted photons.
Our model successfully reproduces the periodic behavior, 
but the actual experimental parameters are found to 
be outside the parameter region 
for the periodic superradiance.
This result implies that some other mechanism(s) is required.
As one example, 
assuming that the field decay rate varies with the electric field, 
the periodic superradiance can be reproduced 
even under the actual experimental condition.
\end{abstract}

\maketitle

%\tableofcontents

%--------------Sec 1--------------------%
\section{Introduction}
\label{sec:introduction}
%--------------Sec 1--------------------%

Superradiance (SR), predicted by Dicke \cite{SR-1954}, 
is one of the coherent phenomena 
caused by cooperative spontaneous emission.
Unlike ordinary spontaneous emission, 
a short intense optical pulse 
with a peak height proportional to $N^{2}$, 
where $N$ is the number of atoms involved, 
is emitted due to the creation of a macroscopic dipole.
For SR generation, 
this macroscopic dipole must develop within a decoherence time $T_{2}$.
Many studies on SR have been carried out 
both theoretically and experimentally.
So far SR has been observed 
in gases \cite{SR-HFgas-1973,cascadeSR-Na-1976,triggeredSR-1980} 
and solid-state materials \cite{SR-O2KCl-1982,SR-nanocrystal-2018,SR-NVcenter-2018,ErYSO-Padova-2020,SR-solid-review-2016}.

Recently, we have observed SR pulses 
with a quasiperiodic time structure 
in an Er:YSO crystal \cite{Hara-pSR}.
The periodic behavior persists 
during a continuous-wave (CW) laser excitation.
We refer to this phenomenon as ``periodic superradiance".
Periodic SR exhibits periodic behavior 
without any external modulation of the input parameters.
This seems similar to the phenomenon 
called as a time crystal \cite{time-crystal-2023}, 
while the origin of periodicity is different.
We provided a qualitative model 
to understand the periodicity of SRs, 
in which a cyclic process of 
a continuous supply of population inversion 
and a sudden burst of SR is repeated.
The excitation power dependence measurements supported 
the validity of the model.
However, our model was entirely qualitative 
and could not reproduce any observed SR quantities, 
such as period, pulse duration, and number of SR photons emitted.

SR is often described by the Maxwell-Bloch equations 
\cite{Benedict-textbook-1996}, 
which are nonlinear differential equations 
representing the dynamics of population, coherence, and radiation. 
For example, a single SR pulse can be obtained 
as a numerical or analytical solution of the Maxwell-Bloch equations 
for a fully excited two-level system.
Because periodic SR has been observed only recently, 
there have been no quantitative theoretical studies 
using the Maxwell-Bloch equations or other frameworks.
A similar phenomenon, self-pulsing in laser media, 
is well known, where persistent pulses are generated 
under CW excitation.
Numerous theoretical studies have addressed this phenomenon 
and successfully reproduced it using rate equations 
\cite{selfpulsing-EDFL-1993}.
However, since coherence plays a central role in SR, 
such treatments cannot be directly applied to periodic SR.

In this paper, 
we report on more quantitative studies of the periodic SR 
than the model we constructed in Ref. \cite{Hara-pSR}.
First, we construct a model based on the Maxwell-Bloch equations 
for a reduced number of energy levels, 
a pair of SR states plus a population reservoir state.
We find that the model, 
abbreviated as the extended two-level Maxwell-Bloch model (X2MB),  
can successfully generate periodic SR pulses 
under CW laser excitation. 
One of the important results 
of the X2MB model is 
the classification of the parameter space into 
two separate regions; 
the periodic SR region and the other region.
Only the periodic SR region exhibits periodic SRs. 
The classification is done 
by analyzing the eigenvalues of the linearized X2MB equations 
around the equilibrium points. 
We find that, somewhat unexpectedly, 
our experimental parameter set lies 
outside the periodic SR region. 
This result holds even when we allow the parameters 
to vary over a wide range of their uncertainties. 
We therefore conclude that some other mechanism(s) is 
required to reproduce the observed periodic SR. 
As an example of such mechanisms, 
we introduce a modulation of the field decay rate, \textit{i.e.}
the rate of electric field emission to the outside of the crystal.
The X2MB model with this modification 
is found to give much more satisfactory results. 
We also derive from the X2MB equations 
a pair of nonlinear differential equations of 
coherence and population difference, 
which are referred to as the truncated two-level Bloch model (T2B). 
We show that this simple two-variable equation contains 
the essential feature of periodic SR pulses. 
In fact, the characteristic quantities of SR, 
such as period, FWHM (full-width-half-maximum) pulse duration 
and number of emitted photons, 
can be expressed analytically by its integral.
We believe that the above studies explore 
intriguing aspects of 
the Maxwell-Bloch equations and shed new light on SR physics.

This paper is organized as follows.
In Sec. \ref{sec:experiment}, 
a brief summary of our experimental results \cite{Hara-pSR} is given.
In Sec. \ref{sec:three-level}, 
the basic equations 
for our experimental system are introduced, 
and they are solved both analytically and numerically.
The necessary condition for the periodic SR is also investigated.
In Sec. \ref{sec:other}, 
we discuss other possible mechanisms that cause the periodic behavior 
and modify the equations with a modulation of the field decay rate.

%--------------Sec 2----- system of ---------%
\section{Summary of experimental results}
\label{sec:experiment}
%--------------Sec 2--------------------%
%

\begin{figure}[t]
\begin{center}
      \includegraphics[width=8.5cm,keepaspectratio]{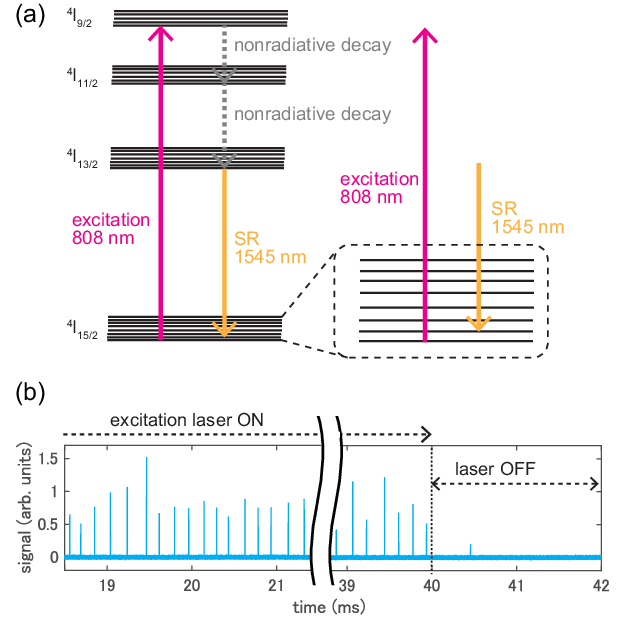}
       \caption{(a) Energy diagram of an Er$^{3+}$ ion doped in a YSO crystal.
       The dashed rounded square shows an enlarged view of 
       the $^{4}$I$_{15/2}$ ground state.
       (b) Example of the waveform of the observed 
       periodic superradiance 
       in the middle 3 ms of the excitation and from 1 ms before the excitation laser is turned off.
       The excitation laser is turned on for 40 ms from $t=0$.
      }
       \label{fig:SRexperiment}
\end{center}
\end{figure}

First, we briefly summarize our experimental results \cite{Hara-pSR}.
The relevant energy diagram of Er$^{3+}$ ions 
doped in a YSO crystal is 
shown in Fig. \ref{fig:SRexperiment} (a).
The crystal field splits each state into multiple Stark levels.
An excitation laser at 808 nm pumps up Er$^{3+}$ ions 
from the ground state 
to the lowest Stark level of the $^{4}$I$_{9/2}$ state.
After rapid decay via nonradiative processes, 
the population accumulates 
in the lowest Stark level of the $^{4}$I$_{13/2}$ state 
with a long lifetime of about 10 ms.
The SR pulses with a wavelength of 1545 nm 
are generated in the transition 
between the two SR states, \textit{i.e.} 
from the lowest Stark level of the $^{4}$I$_{13/2}$ state 
to the second lowest Stark level of the $^{4}$I$_{15/2}$ ground state.
It was confirmed by a wavelength measurement 
using a monochromator.

The Er$^{3+}$ ions at site 2, 
whose number density is about $5 \times 10^{18}$ cm$^{-3}$, 
are used in our experiment.
The excitation laser diameter ($2 w_{0}$) is about 300 $\mu$m 
and the input power is 90 mW.
It propagates through the 6 mm long crystal.
The excitation laser is turned on for 40 ms every 200 ms 
and the 40 ms excitation data are acquired repeatedly.
The detailed setup and procedure are described in Ref. \cite{Hara-pSR}.

Under the above experimental scheme, 
we observed periodic SR pulses 
as shown in Fig. \ref{fig:SRexperiment} (b).
The pulses appear several ms after the excitation laser is turned on ($t=0$), 
and disappear right after it is turned off ($t=40$ ms), 
occasionally followed by one additional SR pulse.
The mean period between SR pulses $(T)$ 
and the standard deviation 
determined by the Gaussian fit to the peak interval histogram 
are 160 $\mu$s and 20 $\mu$s, respectively.
The mean FWHM pulse duration $(\Delta T)$ of SR 
and the standard deviation are 20 ns and 5 ns, respectively.
The number of photons $(N_{\mathrm{SR}})$ 
in a single SR pulse is estimated to be 
$\mathcal{O} (10^{12})$.
For future convenience, we summarize our main experimental results:
\begin{eqnarray}
 && (T,\Delta T, N_{\mathrm{SR}})=(160 \;\mu \textrm{s}, 20 \;\textrm{ns},  10^{12}) .
 \label{eq:main experimental result} 
\end{eqnarray}
Note that in addition to the pulse-to-pulse fluctuations, 
$(T, \Delta T, N_{\mathrm{SR}})$ vary with experimental conditions 
such as the position of the excitation laser; 
therefore, they must be considered 
as representative values \cite{Hara-pSR}.
The periodic time structure can be understood qualitatively 
by a simple model, in which 
a cyclic process of a continuous supply of population inversion 
and a sudden burst of SR emission are repeated.

%--------------Sec 3--------------------%
\section{Extended two-level model based on 
the Maxwell-Bloch equations}
\label{sec:three-level}

In this section, we first construct a model 
based on the Maxwell-Bloch equations with reduced energy levels 
(Sec. \ref{sec:MBeq}). 
This model, which can only be solved by numerical integration, 
is the basis of all the results presented in this paper. 
We refer to this model as an extended two-level model 
based on the Maxwell-Bloch equations 
(X2MB model for short) for a reason explained below.
The X2MB model contains many physical parameters. 
We discuss how we determine 
the central value and uncertainty 
of these parameters (Sec. \ref{sec:parameter}).
They are used in the numerical simulations.
We find that the periodic SR occurs 
only in a certain restricted region of the parameter space. 
The conditions for the parameters necessary 
for the periodic SR to appear 
are clarified 
by analyzing linearized X2MB equations around equilibrium points 
(Sec. \ref{sec:condition}).
We then perform numerical simulations 
with parameters both inside and outside 
of the ``predicted" periodic SR region 
(Sec. \ref{sec:simulation}).
We also present a much simpler model 
that reveals essential features of the periodic SR 
(T2B model) (Sec. \ref{sec:truncated}).
It is a set of nonlinear differential equations 
with only two variables, 
coherence and population difference between the two SR states. 
Analytical solutions of this simple two-variable model 
are shown to reproduce main features of the X2MB simulation results.
Finally we conclude this section 
by searching for a parameter set 
that reproduces our experimental results 
(Sec. \ref{sec:extendedtwolevelsummary}).

%--------------Sec 3--------------------%
\subsection{Extended two-level model: 
concept and basic equations}
\label{sec:MBeq}

As described in the previous section, 
the actual process observed in our experiment 
involves many ionic states and transitions among them.
It is virtually impossible 
to reproduce this complicated process quantum mechanically. 
Fortunately we find that the transitions may be classified 
into two broad categories: 
coherent transitions which require quantum treatments, 
and incoherent ones which can be described by rate equations.
In our case, the transition between the SR states is coherent 
and needs to be treated by the Shr\"{o}dinger (or Bloch) equation. 
All other transitions are incoherent. 
Note that the pumping process by a laser beam is in principle coherent, 
but the Er$^{3+}$ ions pumped up by the laser 
decay rapidly by nonradiative processes and lose their phase memory 
before reaching the upper SR state.
Thus, in fact, we may consider it an incoherent pumping process.

\begin{figure}[t]
\begin{center}
      \includegraphics[width=7cm]{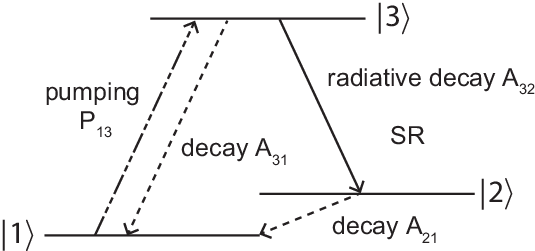}
       \caption{Reduction to an extended two-level system in our model.
       The states $\ket{1}$ and $\ket{2}$ correspond to 
       the lowest and the second lowest Stark levels 
       of the $^{4}$I$_{15/2}$ ground state, respectively.
       The state $\ket{3}$ corresponds to the lowest Stark level 
       of the $^{4}$I$_{13/2}$ state.
       The solid arrow represents the coherent transition.
       The dashed arrows represent the incoherent decay.
       The double chain arrow represents 
       the combination of the pumping (coherent process) and 
       subsequent nonradiative decay (incoherent process).
      }
       \label{fig:threelevel}
\end{center}
\end{figure}

The above considerations leads to a highly simplified model 
of the process shown in  Fig. \ref{fig:threelevel}. 
In the following, this model is explained in more detail 
using Fig. \ref{fig:threelevel}.
The states $\ket{1}$ and $\ket{2}$ correspond to 
the lowest and the second lowest Stark levels 
of the $^{4}$I$_{15/2}$ ground state, 
and $\ket{3}$ corresponds to the lowest level of $^{4}$I$_{13/2}$, 
respectively.
As mentioned above, 
coherence develops only between $\ket{3}$ and $\ket{2}$, 
and generates SR pulses.
This level system can be thought of as a Maxwell-Bloch system 
in the two levels between $\ket{3}$ and $\ket{2}$, 
with the addition of $\ket{1}$ as a reservoir.
Hence the name ``extended two-level model".
The excitation to the lowest Stark level of the $^{4}$I$_{9/2}$ state 
and the subsequent nonradiative decay 
$\ket{3}$ 
are combined and represented 
by a single pumping rate $P_{13}$.
Among the deexcitation processes 
from $\ket{3}$ to the various Stark levels of 
the $^{4}$I$_{15/2}$ ground state, 
only a radiative decay rate $A_{32}$ from $\ket{3}$ to $\ket{2}$ 
contributes to the SR.
The deexcitation processes from $\ket{3}$ 
to other than $\ket{2}$ and 
the other incoherent processes 
between $\ket{3}$ and $\ket{2}$ 
are together represented 
as a deexcitation rate $A_{31}$ 
from $\ket{3}$ to $\ket{1}$.
The population of the state $\ket{2}$ decays to the state $\ket{1}$ 
with a rate of $A_{21}$, 
which is dominated by nonradiative decays, 
making the development of macroscopic dipoles practically impossible.

$\;\\$
We are now in a position to write Maxwell-Bloch type equations 
based on Fig. \ref{fig:threelevel}. 
After applying a rotating-wave approximation and 
a slowly varying envelope approximation \cite{Benedict-textbook-1996}, 
we readily arrive at a set of equations (X2MB equations) shown below: 

\begin{align}
\label{eq:drho11dt}
&\frac{d \rho_{11}}{d t} = -P_{13} \rho_{11} + A_{31} \rho_{33} + A_{21} \rho_{22} , \\
\label{eq:drho22dt}
&\frac{d \rho_{22}}{d t} = \Omega_{s} \rho_{32} + A_{32} \rho_{33} - A_{21} \rho_{22}, \\
\label{eq:drho33dt}
&\frac{d \rho_{33}}{d t} = - \Omega_{s} \rho_{32} + P_{13} \rho_{11} - (A_{31} + A_{32}) \rho_{33}, \\
\label{eq:drho32dt}
&\frac{d \rho_{32}}{d t} = \frac{ \Omega_{s}}{2} ( \rho_{33} - \rho_{22} ) - \gamma_{32} \rho_{32}, \\
\label{eq:Maxwell}
%&\frac{\partial \Omega_{s}}{\partial t} = - \kappa \Omega_{s} + \Omega_{0}^{2} \rho_{32} + \Omega_{0}^{2} R_{\mathrm{sp}} \rho_{33} .
&\frac{\partial \Omega_{s}}{\partial t} = - \kappa \Omega_{s} + \Omega_{0}^{2} \rho_{32} + \Omega_{0} R_{\mathrm{sp}} \rho_{33} .
\end{align}
Here $\rho_{ij}$ are the density matrix elements 
of states $\ket{i}$ and $\ket{j}$.
As shown in Eq. (\ref{eq:drho11dt}), 
the population change of $\ket{1}$ is caused 
by the pumping and decay processes.
The populations of $\ket{2}$ and $\ket{3}$ 
change with the SR generation 
in addition to the pumping and decay processes.
The population changes due to the SR generation are represented 
by the first terms on the right-hand side 
of Eqs. (\ref{eq:drho22dt}) and (\ref{eq:drho33dt}).
The Rabi frequency $\Omega_{s}$ is defined as 
\begin{equation}
        \Omega_{s} \equiv i \frac{d_{32} E_{0}}{\sqrt{3} \hbar} ,
\end{equation}
where $d_{32}$ is the transition dipole moment 
between $\ket{3}$ and $\ket{2}$, 
and $E_{0}$ is the slowly varying envelope of 
the electric field $\vec{E}$ due to the SR radiation.
In this study, without loss of generality, we set 
$\Omega_{s}$ and $\rho_{32}$ 
as real numbers.
Equation (\ref{eq:drho32dt}) represents 
the time evolution of the coherence $\rho_{32}$, 
where $\gamma_{32}$ is the decoherence rate.
As mentioned above, we only consider the coherence 
between these states.
Equation (\ref{eq:Maxwell}) 
is derived from the Maxwell equation, and 
represents the time evolution of the SR electric field.
Note that the differential operator $\partial / {\partial z}$ 
in the Maxwell equation is replaced by $1/L$.
This replacement is justified 
because the crystal length $L$ is short enough that 
the change in $E_0$ can be neglected (homogeneous approximation).  
In this approximation, 
the emission of the SR electric field to the outside of the crystal 
is represented by the field decay rate $\kappa ( = c / n_{0} L)$, 
where $c$ and $n_{0}$ are the speed of light and 
the refractive index of the crystal at a wavelength of SR, respectively.
The second term in Eq. (\ref{eq:Maxwell}) 
represents the field development due to coherence, 
while the last term is a contribution 
to the SR electric field by spontaneous emissions. 
Although the rate $R_{\mathrm{sp}}$ is extremely small, 
it is necessary for the system to start the coherence development. 
In the following we will call $R_{\mathrm{sp}}$ the coherence trigger rate.

$\; \\$
 When analyzing/solving the X2MB model, 
 it is convenient to introduce a frequency 
 that characterizes the time scale of the system.
 This frequency, denoted by $\Omega_{0}$, is given by  
\begin{equation}
    \Omega_{0} = \sqrt{\frac{N_{0} d_{32}^{2} \omega_{32}}{3 \epsilon_{0} \hbar}} , 
\end{equation}
where $N_{0}$ is the number density of Er$^{3+}$ ions at site 2 
and $\omega_{32}$ is the angular frequency 
between $\ket{3}$ and $\ket{2}$.
Another important observation is the hierarchy of time scales in the frequency parameters. 
In fact, 
the relationship among the magnitude of each rate, 
\begin{equation}
\label{eq:hierarchy}
\kappa >> \gamma_{32}, A_{21} >> P_{13}, A_{31}, A_{32} >> R_{\mathrm{sp}} , 
\end{equation}
is useful to understand the properties of our system.
The field decay rate $\kappa$, $\mathcal{O} (10^{10})$ Hz, 
is the highest and dominates the time evolution of the system.
The decoherence rate $\gamma_{32}$, $\mathcal{O} (10^{7 \pm 1})$ Hz 
and the decay rate $A_{21}$, $\mathcal{O} (10^{6})$ Hz, 
are much lower than the field decay rate, 
but are much higher than the other pumping and decay rates 
$P_{13}$, $A_{31}$, and $A_{32}$, $\mathcal{O} (10^{1 \pm 1})$ Hz.
The coherence trigger rate $R_{\mathrm{sp}}$ is 
the smallest among the parameters: 
it does not play a significant role 
once it initiates the coherence development.
The detailed values are given in Table \ref{tab:param}.

%--------------Subsec 3.2--------------------%
\subsection{Experimental parameters}
\label{sec:parameter}
%--------------Subsec 3.2--------------------%

\begin{table*}[t]
\begin{threeparttable}
\caption{Experimental parameters.
The upper part consists of the input parameters 
to the X2MB equations.
The middle part shows the parameters calculated by others.
The bottom part is related to 
the crystal property, energy level, and laser.
The measure of uncertainty of each parameter 
is also displayed.
Here, the uncertainty does not mean $1 \sigma$, 
but is only a guide.
``factor $U$" represents 
the uncertainty from factor $1/U$ to $U$.}
\label{tab:param}
\begin{tabular}{ccccc}
\addlinespace[1.5mm]
\hline
\addlinespace[1.5mm]
symbol & description & central value & measure of uncertainty & remark \\ 
\addlinespace[1.5mm]
\hline
\addlinespace[1.5mm]
$P_{13}$ & pumping rate & 200 Hz\tnote{a} & $\pm 100$ &  \\ 
$A_{32}$ & radiative decay rate & 4 Hz\tnote{a} & factor 10\tnote{b} & \\ 
$A_{31}$ & decay rate & 100 Hz\tnote{a} & $70 \leq A_{31} \leq 110$ &  \\
$A_{21}$ & decay rate & $2.5 \times 10^{6}$ Hz & factor 3 & Ref. \cite{Hara-pSR} \\
$\gamma_{32}$ & decoherence rate & $10^{7}$ Hz\tnote{a} & factor 10\tnote{a} &  \\
$R_{\mathrm{sp}}$ & coherence trigger rate & $10^{-30} \times \Omega_{0}$\tnote{a} & see main text &  \\ 
%R& to SR & & estimate\tnote{a} & \\
\addlinespace[1.5mm]
\hline
\addlinespace[1.5mm]
$\kappa$ & field decay rate & $2.8 \times 10^{10}$ Hz & factor 3\tnote{c} & $c / n_{0} L$\\
$d_{32}$ & transition dipole moment & $2.3 \times 10^{-32}$ C m & factor 3 & $\sqrt{\frac{3 \pi \epsilon_{0} \hbar c^{3} A_{32}}{\omega_{32}^{3}}}$ \\
$\Omega_{0}$ & characteristic frequency of SR & $3.3\times10^{10}$ rad/s & factor 10 &  $\sqrt{\frac{N_{0} d_{32}^{2} \omega_{32}}{3 \epsilon_{0} \hbar}}$ \\
\addlinespace[1.5mm]
\hline
\addlinespace[1.5mm]
$N_{0}$ & number density of Er$^{3+}$ ions at site 2 & $4.7 \times 10^{18}$ cm$^{-3}$ & factor 10\tnote{d} &  0.1 \% concentration \\
$N_{\mathrm{Er}}$ & total number of Er$^{3+}$ ions at site 2 & $2.0 \times 10^{15}$ & factor 10 & $N_{0} \pi w_{0}^{2} L$  \\
$n_{0}$ & refractive index & 1.8 & $\pm 0.1$ & Ref. \cite{petersen-PhD-2011,YSO-microwave-2015}\tnote{e} \\
$L$ & crystal length & 6 mm & negligible & \\
$w_{0}$ & excitation laser radius & 150 $\mu$m & $\pm 10$ & \\
$\omega_{32}$ & angular frequency & $1.2 \times 10^{15}$ rad/s & negligible & Ref. \cite{ErYSO-spectroscopy-2006} \\
\addlinespace[1.5mm]
\hline
\end{tabular}
\begin{tablenotes}
\item[a] Discussed in the main text.
\item[b] This uncertainty is mainly due to the uncertainty in $N_{0}$.
%\item[c] The number is rounded. 
\item[c] The uncertainty determined from the crystal length is small, but the actual uncertainty 
should be large.
\item[d] The uncertainty determined from the concentration of the preparation is small, but the actual uncertainty is large due to the inhomogeneity of the crystal.
\item[e] The value of the refractive index at room temperature is estimated using the Sellmeier equation \cite{petersen-PhD-2011}. Applied from the microwave case \cite{YSO-microwave-2015}, the refractive index is only a few percent different between 4 K and room temperature.
\end{tablenotes}
\end{threeparttable}
\end{table*}

In the following part, 
we consider the condition for the periodic SR 
and perform the numerical simulation 
for the different experimental parameter sets.
Before that, we introduce the experimental parameters.
The parameters in our experiment are shown in Table \ref{tab:param}.
The upper part consists of the input parameters 
necessary to solve the X2MB equations.
Only these parameters will be discussed in detail here.
The pumping rate $P_{13}$ is estimated 
from the approximately 85 \% absorption of the 90 mW excitation laser.
The radiative decay rate $A_{32}$ is estimated 
from the fluorescence spectrum \cite{ErYSO-spectroscopy-2006} 
and the absorption spectrum measured in our system.
The decay rate $A_{31}$ is estimated 
by subtracting $A_{32}$ from the decay rate 
determined by the fluorescence lifetime of $\ket{3}$ 
($9.20 \pm 0.02 $ ms \cite{ErYSO-spectroscopy-2006}).
The decay rate $A_{21}$ is measured 
by the time constant of the change in absorption 
from the lower state of SR to the third lowest Stark level of 
the $^{4}$I$_{9/2}$ state \cite{Hara-pSR}.
The decoherence rate $\gamma_{32}$ is estimated 
from the fact that the decoherence time should be 
longer than the duration of the SR pulses ($\mathcal{O} (10)$ ns) 
and shorter than twice the lifetime of the lower state of SR 
($2 \times 400$ ns).
In principle, the coherence trigger rate $R_{\mathrm{sp}}$ may be estimated by considering the spontaneous emission rate, 
its phase variation, and the modes of SR emission. 
In practice, however, it is difficult to estimate accurately 
because of its random nature. 
Fortunately, 
the final result is insensitive to this parameter. 
For example, even if it differs by ten orders of magnitude, 
the results for the main quantities 
- period, FWHM pulse duration, and SR photon number - 
as well as the timing of the first SR pulse, 
change by a factor of less than 2.
We arbitrarily assume a very small value in this paper; $R_{\mathrm{sp}}=10^{-30} \times \Omega_{0}$.

A few comments are in order here.
First, we note that most of the parameters are not 
accurately determined. 
We give rough estimates of their uncertainties 
in Table \ref{tab:param} 
and in its footnotes, along with their central values.
Second, given the rather large parameter uncertainties,  
we varied them widely in the simulation to study their dependence.
We find that our conclusion is essentially unchanged, as shown below. 

%--------------Subsec 3.3--------------------%

\subsection{Conditions for periodic SR} 
\label{sec:condition}

Given a set of parameters, 
it is desirable to know whether or not the X2MB equations 
produce periodic SRs. 
Since it is impossible to run a simulation for every parameter set, 
such predictability is essential 
to draw the conclusion of this section. 
To this end, we linearize the X2MB equations 
around equilibrium points, 
and analyze the eigenvalues of the resulting Jacobi matrix. 
We classify the parameter space based on the sign of the eigenvalues 
and exclude the region where periodic SR cannot occur.
This analysis is elaborated below.

\subsubsection{Linearized X2MB equations and Jacobi matrix}
\label{par:Jacobi maxtrix}

We first rewrite the original X2MB equations, 
Eqs. (\ref{eq:drho11dt}) - (\ref{eq:Maxwell}), 
into dimensionless forms. 
This is done in Appendix \ref{app:derivation} 
to obtain Eqs. (\ref{eq:du0dttilde}) - (\ref{eq:du3dttilde}). 
We then expand them around the equilibrium points 
given by Eqs. (\ref{eq:u0e}) - (\ref{eq:u3e}). 
The final form of the linearized equations is 
\begin{equation}
\label{eq:linearized}
\frac{d \vec{y}}{d \tilde{t}} = J \vec{y}, 
\qquad J= \begin{pmatrix}
    -b_{00} & 1 & 0 & 0  \\
    u_{2}^{e} & -b_{11} & u_{0}^{e} & 0 \\
    - u_{1}^{e} & - u_{0}^{e} & b_{22} & b_{23} \\
    0 & 0 & b_{32} & b_{33} \\
    \end{pmatrix} , 
\end{equation}
where $J$ is a Jacobi matrix 
and $y_{i} \equiv u_{i} - u_{i}^{e} (i=0,1,2,3)$ 
is a deviation from the equilibrium point.
The solution of Eq. (\ref{eq:linearized}) can be expressed as 
$\vec{y} = \sum_{i=0}^{3} C_{i} \vec{x}_{i} \exp{(\lambda_{i} \tilde{t})}$ 
using the eigenvalues $\lambda_{i}$ of $J$, 
the eigenvectors $\vec{x}_{i}$ for each eigenvalue, 
and the constants $C_{i}$.
As noted above, the behavior around the equilibrium point 
is determined by the eigenvalues of the Jacobi matrix. 
For example, if the real parts of the eigenvalues are all negative, 
then the system approaches the equilibrium point 
($\vec{y} \to \vec{0}$). 
In other words, the periodic SR generation requires 
the system to be outside such a region.
In fact, there are two equilibrium solutions to the X2MB equations.
One is the ``trivial solution" with zero electric field, 
and the other is the ``nontrivial solution" with finite electric field.
For periodic SR, the system is expected to move away 
from the equilibrium point of the nontrivial solution.
Thus, the system should be in the region 
where at least one real part of the eigenvalues is positive 
for the nontrivial equilibrium point.
This is a necessary condition, but it may not be sufficient.
It does not hold in principle, 
but has been investigated under several conditions 
and found to be correct.
Our expectation is confirmed by the numerical simulation results.
We refer to the above region, 
where the periodic SR is expected to occur, 
as the ``periodic SR region".

\subsubsection{Plot of the periodic SR region}
\label{par:asymptotic stable region}

\begin{figure}[t]
\begin{center}
     \includegraphics[width=8.5cm]{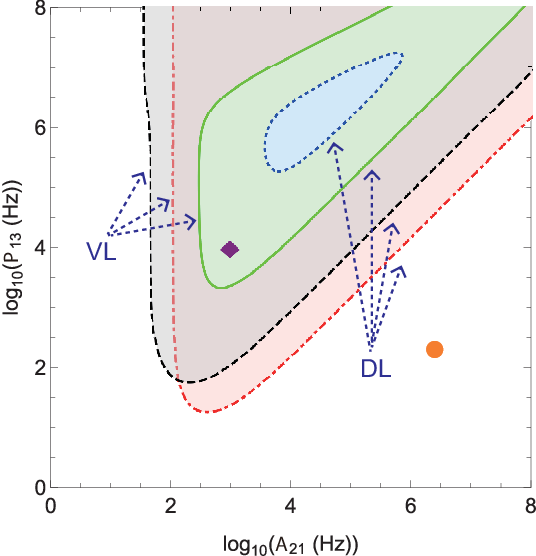}
       \caption{Periodic SR regions 
       in the $(A_{21},P_{13})$ plane for different $A_{32}$ 
       with $(A_{31},\gamma_{32})$ fixed at 
       $(\underline{10^{2}}, \underline{10^{7}})$ Hz.
       The periodic SR regions are colored.
       The blue dotted, green solid, black dashed, 
       and red dot-dashed lines are the region boundaries 
       for $A_{32}$ = 3, $\underline{4}$, 20, and 100 Hz, respectively.
       VL; vertical line.
       DL; diagonal line.
       The purple diamond represents 
       the parameter used in the simulation 
       of Fig. \ref{fig:simulation} (a).
       The orange circle represents 
       the actual experimental parameter.}
       \label{fig:stability}
\end{center}
\end{figure}

As pointed out in Sec.\ref{sec:parameter}, 
there are five important parameters in our model; 
$P_{13}$, $A_{31}$, $A_{32}$, $A_{21}$, and $\gamma_{32}$. 
Among them, only $P_{13}$ is controllable by the experiment. 
Nevertheless, we have varied all these parameters 
to study the parameter dependence of the periodic SR region. 
For easier understanding, 
only $(A_{21},P_{13})$ are treated as variables at first, 
while the other parameters are fixed at certain values. 
Then all other parameters are varied one by one, 
as is discussed below.

Figure \ref{fig:stability} shows the periodic SR regions 
in the $(A_{21},P_{13})$ plane for different $A_{32}$ 
with $(A_{31}, \gamma_{32})$ fixed at 
$(\underline{10^{2}},\underline{10^{7}})$ Hz.
Here, and below, the underline below the numerical value 
indicates that it corresponds to the actual experimental value 
shown in Table \ref{tab:param}.
The blue dotted, green solid, black dashed, 
and red dot-dashed lines are the boundaries of the periodic SR regions 
for $A_{32}$ = 3, $\underline{4}$, 20, and 100 Hz, respectively.
The periodic SR regions within these boundaries are colored.
The green region corresponds to the case with the actual value of 
$(A_{31}, A_{32}, \gamma_{32})$, but not of $(A_{21}, P_{13})$.
When $A_{32}$ is varied to the higher frequency, 
the diagonal line (DL) shifts to the lower right.
The vertical line (VL) shifts to the lower frequency side 
for $A_{32} < 20$ Hz, 
and it shifts to the higher frequency side for $A_{32} > 20$ Hz.
We also examine how the results change 
when the other parameters, $\gamma_{32}$ and $A_{31}$, are varied.
If $\gamma_{32}$ is changed from $0$ to $10^{8}$ Hz 
with $(A_{31},A_{32}) = (\underline{10^{2}},\underline{4})$ Hz, 
the results are almost the same as the green region. 
If $A_{31}$ is changed from $10^{2}$ Hz to 10 or $10^{3}$ Hz 
with $(\gamma_{32}, A_{32}) = (\underline{10^{7}},\underline{4})$ Hz, 
the vertical line (VL) shifts to left or right.
In the next section, we examine the validity of 
the periodic/non-periodic classification 
discussed above using simulation.

\subsection{Numerical simulation} 
\label{sec:simulation}

\subsubsection{Results of the numerical simulation}
\label{sec:result_sim}

\begin{figure*}[t]
\begin{center}
      \includegraphics[width=17cm]{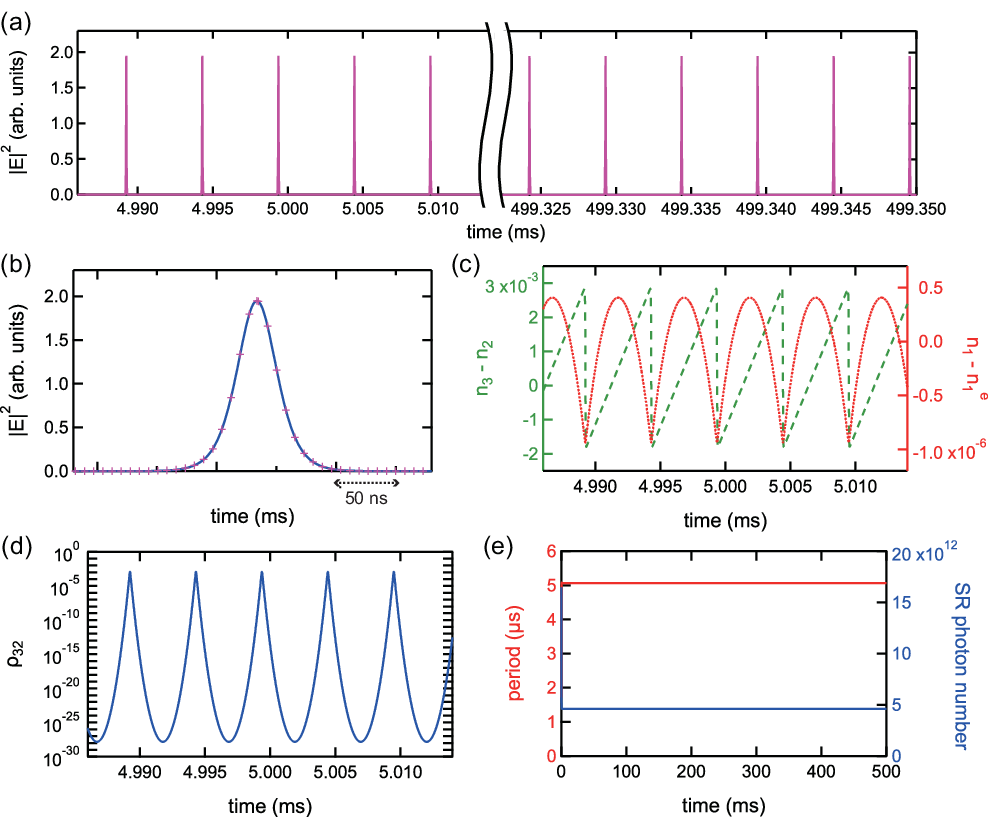}
       \caption{(a) Example of a numerical simulation result 
       within the periodic SR region.
       The waveforms 
       at initial part ($t \sim 5$ ms) 
       and after a very long time ($t \sim 500$ ms) are shown.
       The parameters are $(P_{13},A_{21},A_{31},A_{32},\gamma_{32})=(10^{4},10^{3},\underline{10^{2}},\underline{4},\underline{10^{7}})$ Hz.
       The magenta line is the square of the absolute value of 
       the electric field, which is proportional to 
       the instantaneous light intensity due to the emitted photons.
       (b) Extended view around a single SR pulse.
       The magenta crosses are the simulation results.
       The blue line is the fit by a sech-squared function.
       (c) Time evolution of the population inversion 
       between $\ket{3}$ and $\ket{2}$ 
       (left axis, green dashed line) 
       and the population of the ground state $n_{1}$ 
       with respect to the equilibrium value 
       $n_{1}^{e} (\sim 0.05)$ 
       (right axis, red dotted line).
       (d) Time evolution of the coherence 
       between $\ket{3}$ and $\ket{2}$ on a logarithmic scale.
       (e) Period (left axis, red line) 
       and SR photon number (right axis, blue line) 
       until $t = 500$ ms.}
       \label{fig:simulation}
\end{center}
\end{figure*}

In this section, we present numerically integrated solutions 
of the X2MB equations, 
actually the dimensionless version 
Eqs. (\ref{eq:drho11dttilde}) - (\ref{eq:depsilonSRdttilde}), 
solved by the Runge-Kutta-fourth-order method. 
We first show a typical solution 
both inside and outside the periodic SR regions.
We then give a brief summary of our numerical simulation studies.

\paragraph{Inside the periodic SR region}
\label{par:unstable}

As an example, we arbitrarily pick up 
a parameter set 
$(P_{13},A_{21},A_{31},A_{32},\gamma_{32})=(10^{4},10^{3},\underline{10^{2}},\underline{4},\underline{10^{7}})$ Hz 
within the periodic SR region.
It is represented by the purple diamond 
in Fig. \ref{fig:stability}.
The simulation is started at $t = 0$ with the initial populations 
$\rho_{11}$, $\rho_{22}$, and $\rho_{33}$ 
determined from the Boltzmann distribution at 4 K.
This numerical simulation is run up to $t=500$ ms.
Figure \ref{fig:simulation} (a) shows the square of the electric field of both initial ($t \sim$ 5 ms) 
and final ($t \sim$ 500 ms) stages, 
which corresponds to the instantaneous light intensity 
of the emitted photons.
In the simulation, the pulses start to appear 
at around $t\sim1$ $\mu$s; 
however, this initial stage is not shown 
in Fig. \ref{fig:simulation} (a).
The first SR pulse emerges after a finite delay time 
($\sim 1$ $\mu$s in the present case), 
reflecting both the coherence buildup 
and the formation of a population inversion through pumping.
The peak interval and pulse shape gradually change 
until about $t=5$ ms, 
after which they remain constant, 
in other words, the optical pulses of the identical shape 
are generated at the constant interval.
The optical pulses are generated periodically 
at least until $t=500$ ms.
There is no tendency for the SR behavior to change.
One of the pulses at around $t = 5$ ms 
is shown in Fig. \ref{fig:simulation} (b).
The magenta crosses are the simulation output.
The FWHM pulse duration is $\mathcal{O} (10)$ ns, 
which is $10^6$ times shorter than the lifetime of 
the higher state of this transition (10 ms).
The short pulse duration is one of the characteristics of SR.
The blue line is the fit by a sech-squared function, 
which is another characteristics of SR 
when the macroscopic polarization (coherence) develops 
homogeneously over the target \cite{pure-oscillatory-SF-1975}.
Figure \ref{fig:simulation} (c) shows 
the population inversion between $\ket{2}$ and $\ket{3}$ 
(left axis, green dashed line) 
and the population of the ground state $n_{1}$ 
with respect to the equilibrium value $n_{1}^{e}$ 
(right axis, red dotted line).
The population inversion increases linearly 
due to the pumping and the various decay processes, 
and decreases suddenly when the SR pulses are generated.
The number of photons emitted in a single SR pulse 
is proportional to the sudden decrease of the population inversion.
The time evolution of the coherence $\rho_{32}$ 
is shown in Fig. \ref{fig:simulation} (d) on a logarithmic scale.
After an SR burst, the coherence initially decreases, 
mainly due to a change in the sign of the population inversion, 
as described by Eq. (\ref{eq:drho32dt}).
The minimum value of $\rho_{32}$ 
is close to the order of $R_{\mathrm{sp}}$, 
as revealed by numerical simulations.
As constant pumping gradually restores a population inversion, 
$\rho_{32}$ grows again, eventually leading to the next SR burst.
This development of coherence under constant pumping 
results in periodic SR, in agreement with our expectations.
Amplified spontaneous emission (ASE) 
can also occur in the presence of population inversion 
without coherence.
It was observed in our experiment, 
but its contribution is much smaller than that of SR, 
and we do not discuss it further in this paper.
Similar periodic behavior can be reproduced 
for other parameters within the periodic SR region.
While on the other hand, 
the change of the ground state population ($n_{1}$) 
is much smaller than that of the population 
inversion between $\ket{2}$ and $\ket{3}$ ($n_{3} - n_{2}$).
This result corresponds to Eq. (\ref{eq:du3dtau}) in Appendix \ref{app:derivation}.
Figure \ref{fig:simulation} (e) shows 
the period (left axis, red line) 
and SR photon number (right axis, blue line) 
until $t=500$ ms.
After about 5 ms of the start of excitation, 
the periodic behavior does not change.
Although about $10^{5}$ SRs are generated at $t = 5 \sim 500$ ms, 
the relative standard deviations of the period and photon number 
are less than $10^{-4}$, 
and it can be argued that the SR behavior does not change 
at least until $t=500$ ms.
The main purpose of the above simulation 
is to show that periodic SR is reproduced 
without any external periodic input 
in a certain parameter space.
The period differs significantly from the experimental value 
due to the different parameters from the experimental values.

Periodic SR we are considering is a phenomenon 
that differs from relaxation oscillation in laser systems.
Coherence plays an important role in our system.
The rate equations used in the case of relaxation oscillation 
only gives a damped oscillation solution.
See Appendix \ref{app:rateeq} for more details.

\paragraph{Outside the periodic SR region}
\label{par:stable}

\begin{figure}[t]
\begin{center}
       \includegraphics[width=9cm]{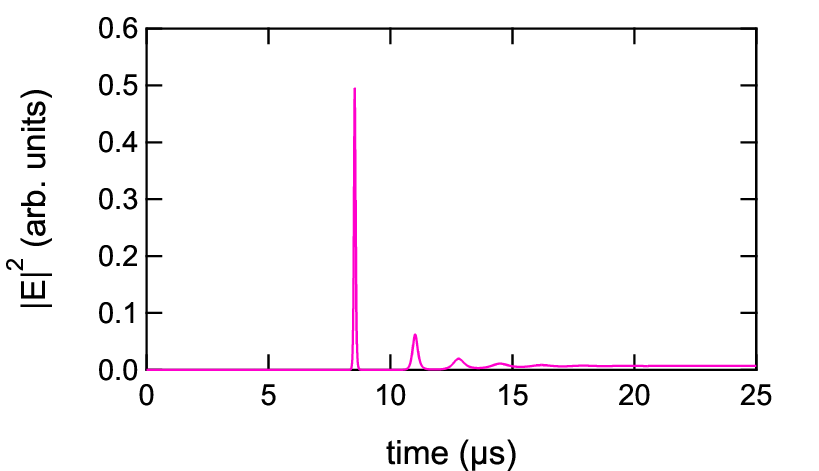}
       \caption{Example of a numerical simulation result 
       outside the periodic SR region.
       The parameters are $(P_{13},A_{21},A_{31},A_{32},\gamma_{32})=(\underline{200},\underline{2.5\times10^{6}},\underline{10^{2}},\underline{4},\underline{10^{7}})$ Hz, 
       which are the actual experimental parameters.
      }
       \label{fig:ESRdecay}
\end{center}
\end{figure}

Outside the periodic SR region, 
we pick up the parameter set 
$(P_{13},A_{21},A_{31},A_{32},\gamma_{32}) = (\underline{200},\underline{2.5\times10^{6}},\underline{10^{2}},\underline{4},\underline{10^{7}})$ Hz.
It corresponds to the actual experimental parameters 
represented by the orange circle in Fig. \ref{fig:stability}.
The initial populations are the same as above.
Figure \ref{fig:ESRdecay} shows 
the instantaneous light intensity due to the emitted photons.
Several optical pulses are generated, 
but they decay within 15 $\mu$s 
after the start of the pumping.
The system reaches equilibrium, 
and the variables 
such as population, coherence, and electric field 
are almost constant.
Thus, the periodic SR is no longer observable.
This behavior is consistent with our expectation in Sec. \ref{sec:condition}.
Similar decays are observed 
for other parameters outside the periodic SR region.

\subsubsection{Brief summary of numerical simulation}
\label{sec:summary_sim}

As shown above, 
we performed the numerical simulation of the X2MB model 
by selecting one point each from 
inside and outside the periodic SR region.
As expected from the analysis of the eigenvalues 
of the linearized X2MB equation, 
the periodic SR is reproduced inside the periodic SR region 
and the SR pulses decay outside the region.
Furthermore, we confirmed that this conclusion holds 
for at least 20 other points within each region.
Near the boundary, however, 
the behavior of the numerical simulation 
does not exactly match the result of 
the eigenvalue analysis; 
roughly speaking, the boundary is blurred over a width of 
$\sim A_{21}/4$.

\subsection{Truncated two-level Bloch model}
\label{sec:truncated}

As mentioned in the introduction to this section, 
a much simpler model, 
the truncated two-level Bloch model (T2B model), 
can be derived from the X2MB equations.
The model, 
which captures the essential physics of cyclic motion, 
can be intuitively understood as follows: 
it assumes only coherence and constant pumping terms 
between the two SR levels of $\ket{2}$ and $\ket{3}$.
The coherence grows as the upper-state population increases 
until it suddenly emits SR pulses.
The population damped in the lower state 
will be pumped up again to repeat the cycle.
The model ignores 
all spontaneous decay and decoherence processes 
that are considered to be in a steady state, 
leaving no contribution to differential equations.
The T2B model is useful to shed light on essential features 
of the SR periodicity and will be discussed in more detail 
in Appendix A.

To arrive at the T2B model, we first eliminate $\rho_{11}$ 
using the relation $\rho_{11}+\rho_{22}+\rho_{33}=1$.
Second, the time derivative in Eq. (\ref{eq:Maxwell}) is ignored 
because the field decay rate $\kappa$ is very high 
and dominates the time evolution of the system.
This approximation makes the SR electric field 
directly proportional to the coherence $\rho_{32}$.
The tiny contribution of 
$R_{\mathrm{sp}}$ is also ignored.
Finally, we assume 
the population of $\ket{1}$ is nearly empty 
based on the physics picture described in Appendix A.
With these approximations and 
after making 
Eqs. (\ref{eq:drho11dt}) - (\ref{eq:Maxwell}) dimensionless, 
the following differential equations are obtained: 
\begin{align}
\label{eq:dXdtau}
&\frac{d X}{d \tau} = X Y, \\
\label{eq:dYdtau}
&\frac{d Y}{d \tau} = - ( X^{2} - 1) . 
\end{align}
Here $X$ is proportional to the coherence $\rho_{32}$, 
and thus to the SR electric field.
$Y$ is proportional to the deviation of $\rho_{33} - \rho_{22}$ 
from the equilibrium point.
$\tau$ is the dimensionless time.
In this paper, we refer to these equations as the T2B model.

\begin{figure}[t!]
\begin{center}
       \includegraphics[width=7.5cm]{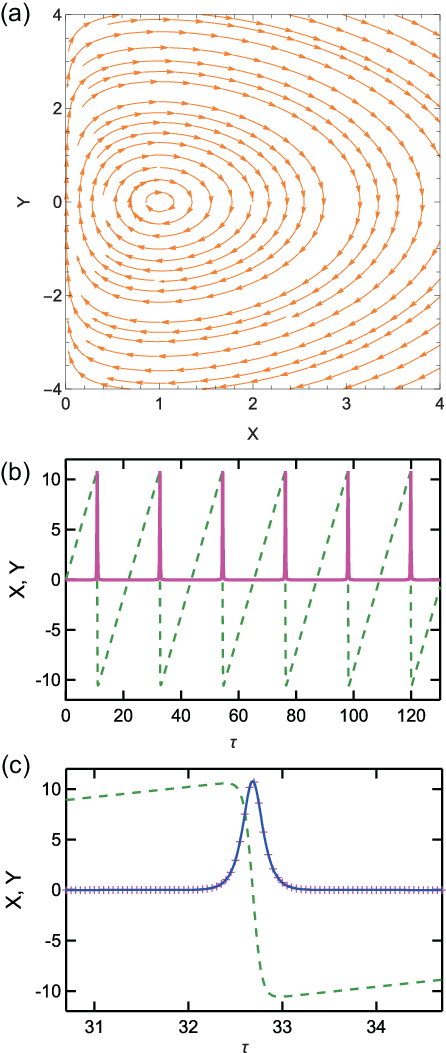}
       \caption{(a) Streamline of Eqs. (\ref{eq:dXdtau}) and (\ref{eq:dYdtau}) in the XY-plane.
       (b) Solution of the T2B model.
       The magenta solid line is $X$, 
       which is proportional to the SR electric field.
       The green dashed line is $Y$, 
       which is proportional to 
       the deviation of the population inversion 
       from the equilibrium point.
       The minimum coherence required to find $C$ 
       is assumed to be $\mathcal{O} (10^{-29})$.
       (c) Extended view of (b) around a single pulse.
       The magenta crosses are $X$.
       The blue line is the fit to $X$ by a sech function.
       The green dashed line is $Y$.
      }
       \label{fig:analytical}
\end{center}
\end{figure}

Let us examine what the T2B model tells us. 
As is well known \cite{strogatz-2001-nonlinear}, 
the global feature is best analyzed by a streamline plot. 
Figure \ref{fig:analytical} (a) shows the streamlines of 
Eqs. (\ref{eq:dXdtau})-(\ref{eq:dYdtau}).
As can be seen, the stream flows clockwise and circles 
around the equilibrium point, (1,0), 
in the XY plane, suggesting cyclic behavior.
The line along which the solution $(X,Y)$ moves depends on 
the initial value of $(X,Y)$.
The cyclic motion in the XY plane corresponds to the periodic behavior.

Figure \ref{fig:analytical} (b) shows 
the numerical solution of the T2B model.
The magenta solid and green dashed lines 
are $X$ and $Y$, respectively.
Since $X$ is proportional to the SR electric field 
in our approximation, 
the magenta solid line indicates 
that the short optical pulses are generated periodically.
The green dashed line indicates 
that the population inversion increases linearly 
due to the pumping and decay processes 
and that it drops rapidly due to the SR pulse generation.
Figure \ref{fig:analytical} (c) 
is the extended view of Fig. \ref{fig:analytical} (b) 
around a single pulse.
The magenta crosses are $X$.
The blue line is the fit to $X$ by a sech function, 
which is one of the characteristics of SR 
when coherence develops homogeneously 
over the target \cite{pure-oscillatory-SF-1975}.

The analytical solution of the period $T$ is obtained as follows.
A simple manipulation of 
Eqs. (\ref{eq:dXdtau}) and (\ref{eq:dYdtau}) yields 
\begin{equation}
\label{eq:dXdY}
 Y dY = \frac{1-X^{2}}{X} dX .
\end{equation}
Integrating Eq. (\ref{eq:dXdY}), we obtain 
\begin{equation}
\label{eq:Y2}
Y^{2} = C - (X^{2} - \ln (X^{2})) , 
\end{equation}
where the integration constant $C$ has the relationship 
$C = \ln (e^{X_{\mathrm{max,min}}^{2}}/X_{\mathrm{max,min}}^{2}) = 1 + Y_{\mathrm{max,min}}^2$ 
with the maximum and minimum values of $X$ and $Y$.
After eliminating $Y$ using Eq. (\ref{eq:Y2}), 
Eq. (\ref{eq:dXdtau}) becomes 
\begin{equation}
\label{eq:dtau}
d \tau = \Bigl\{ X \sqrt{C - \ln (e^{X^{2}}/X^{2})} \Bigr\}^{-1} d X . 
\end{equation}
Integrating Eq. (\ref{eq:dtau}) along the streamline 
from $X_\mathrm{min}$ to $X_\mathrm{max}$, 
the period $T$ is obtained as follows.
\begin{align}
\label{eq:period}
&T = \int_{w_{\mathrm{min}}}^{w_{\mathrm{max}}} dw \Bigl\{ w \sqrt{C - \ln (e^{w}/w)} \Bigr\}^{-1}, \\
& \hspace{5mm} w_{\mathrm{min}}=-W_0(-e^{-C}), 
    \qquad w_{\mathrm{max}}=-W_{-1}(-e^{-C}) ,
\end{align}
where $w=X^{2}$ and 
$W_{J}$ is the Lambert $W$ function with branch index $J$.

Similarly, 
the half of the FWHM pulse duration $\Delta T / 2$ is obtained 
by integrating Eq. (\ref{eq:dtau}) along the streamline 
from $X_{\mathrm{max}}/\sqrt{2}$ to $X_{\mathrm{max}}$, 
recalling that $X$ is proportional to the SR electric field.
The FWHM pulse duration $\Delta T$ is given by 
\begin{equation}
\label{eq:FWHM}
\Delta T = \int_{w_{\mathrm{max}} / 2}^{w_{\mathrm{max}}} dw \Bigl\{ w \sqrt{C - \ln (e^{w}/w)} \Bigr\}^{-1}. 
\end{equation}

By the way, by expanding $W_{J} (x)$ 
in series of $x$ or $\ln (-x)$ \cite{corless-1996-lambert}, 
we can derive approximate fomulae for $T$ and $\Delta T$ 
when $C \gg 1$: 
$\displaystyle T \simeq ( 2 / \sqrt{C} ) \ln 2\sqrt{C}+ 2\sqrt{C}$ 
and $\Delta T\simeq 1.76/\sqrt{C}$.

The number of photons emitted in a single SR pulse 
is estimated as $Y_{\mathrm{max}} - Y_{\mathrm{min}}$ 
multiplied by an appropriate factor, 
since the sudden drop in the population difference 
in Fig. \ref{fig:analytical} (b) is caused by the SR generation.

The T2B model succeeds in capturing 
the essential features of SR periodicity.
First, the periodic pulses can be reproduced by the T2B model.
Second, the T2B model can also reproduce 
the temporal shape of the electric field 
of the sech function, which is characteristic of SR.
Furthermore, the period, the FWHM pulse duration, 
and the number of photons emitted 
are expressed analytically with a single integration constant $C$. 
Unfortunately, it is not yet possible to derive $C$ 
from a given set of parameters.

Finally, the validity of the analytical solution 
of the T2B model is mentioned.
Only a summary is given here, 
see Appendix \ref{sec:comparison} for more details.
We compare the analytical solution of the T2B model 
and the numerical simulation result of the X2MB model 
for the same parameter set.
The period, the FWHM pulse duration, and the number of emitted photons 
are in good agreement.
This result shows the validity of the T2B model 
derived using several approximations.

\subsection{Summary of the extended two-level model}
\label{sec:extendedtwolevelsummary}

In this section, we have constructed a model 
to reproduce the observed periodic SRs. 
The model, X2MB, is based on the Maxwell-Bloch equations 
with the reduced number of energy levels, 
and can be solved by numerical integrations.
We first found that the X2MB results show periodic SRs for constant input parameters in a certain parameter region.
By analyzing the eigenvalues of the linearized X2MB equations, 
we were able to successfully identify the parameter space 
where periodic SRs occur. 
Detailed numerical simulations support 
the validity of the eigenvalue analysis.
We also derived the T2B model from X2MB.
It is at least partially integrable and gives analytical expressions 
for the SR period $T$, pulse duration $\Delta T$ 
and the number of SR photons. 
These predictions are found to be in good agreement 
with the numerical simulation results of the X2MB model 
(see Appendix \ref{sec:comparison}).

The model indicates 
that self-organizing periodicity stems from 
the nonlinear nature of the differential equation.
We also find that coherence 
is another important factor in this respect. 
Standard rate equations 
that use stimulated emission/absorption terms 
do not produce everlasting periodic pulses 
in the desired parameter space 
(see Appendix \ref{app:rateeq} for details).

The important question here is whether 
the model as a whole can reproduce our experimental results.
As already mentioned, the result of the simulation 
with the actual experimental parameters 
does not show any periodicity.
In other words, the experimentally realized periodic SR is 
outside the theoretically predicted periodic SR region.
This conclusion holds even if 
we consider the uncertainties of the parameters. 
The above conclusion implies that some other mechanism is needed 
to reproduce the observed periodic SR 
for the actual experimental parameter set. 

%--------------Sec 4--------------------%
\section{Other possible mechanisms for periodic SR}
\label{sec:other}

%--------------Sec 4--------------------%

In this section, we present other possible mechanisms 
that may explain the experimental results.
So far, we have treated the parameters in X2MB 
as constants in time in a single simulation run.
In the following, we assume that the parameters may vary dynamically 
due to certain nonlinear effects.
For example, the decoherence rate $\gamma_{32}$ and 
the decay rates $A_{31}$ and $A_{21}$, 
associated with phonon decay processes, 
may vary due to the local increase in crystal temperature 
caused by the electric field generated.
Recalling the optical Kerr effect, 
where the refractive index varies 
with the electric field intensity, 
the field decay rate $\kappa$ may also vary 
due to the electric field.
The intensity-dependent loss rate was also discussed 
in the previous study on self-pulsing \cite{shang-selfpulsing-2020}.
Furthermore, the reduction of the group velocity 
in a gain medium, as reported in Ref. \cite{slowlight-highgain-1971}, 
provides another example of dynamical variation 
in the field decay rate $\kappa$.

%--------------Subsec 4.1--------------------%
\subsection{Field decay rate modulation}
\label{subsec:kappamod}
%--------------Subsec 4.1--------------------%

In this section, 
we study in more detail the case 
in which the field decay rate $\kappa$ depends on 
the intensity of the SR electric field ($\kappa$ modulation).
The origin of the modulation can be intuitively understood 
from the following physical picture. 
When an electric field begins to grow, 
a standing wave is formed in the crystal 
and generates a strong-weak-intensity stripe.
This leads to a periodic variation of the refractive index $n$ 
due to the optical Kerr effect.
Then the modulation of $n$ increases 
the effective reflection coefficient at the media boundaries 
(\textit{i.e.} edges with different $n$),  
making an apparent cavity (or crystal) length longer 
and thus $\kappa$ smaller. 
A more rigorous derivation, along with a numerical estimate, 
can be found in Appendix \ref{app:mechanism}.

We assume that the field decay rate $\kappa$ 
in Eq. (\ref{eq:Maxwell}) decreases 
for the strong SR electric field.
For simplicity, we binarize $\kappa$ 
with respect to the threshold electric field 
as follows.
\begin{equation}
\label{eq:kappa}
    \kappa (E_{0}) = \left\{
    \begin{array}{ll}
       \kappa_{0}  &  (|E_{0}| < E_{\mathrm{th}}) \\
       \kappa_{0}/q & (|E_{0}| > E_{\mathrm{th}}), 
    \end{array}
    \right.
\end{equation}
where $E_{0}$ is the slowly varying envelope of the SR electric field, 
$\kappa_{0}=c/n_{0} L$ is the original field decay rate, 
and $E_{\mathrm{th}}$ is the electric field threshold.
The quantity $q$ $(q>1)$ defines how much $\kappa$ decreases 
and can approximately be regarded as the finesse of the cavity.
In the following, we will refer to $(q,E_{\mathrm{th}})$ 
as the modulation parameters.
We note that 
binarising $\kappa$ (Eq. \eqref{eq:kappa}) 
is merely a convenience to simplify the calculation.
See Appendix \ref{app:mechanism} for more detail.

\begin{figure}[t]
\begin{center}
      \includegraphics[width=8.5cm]{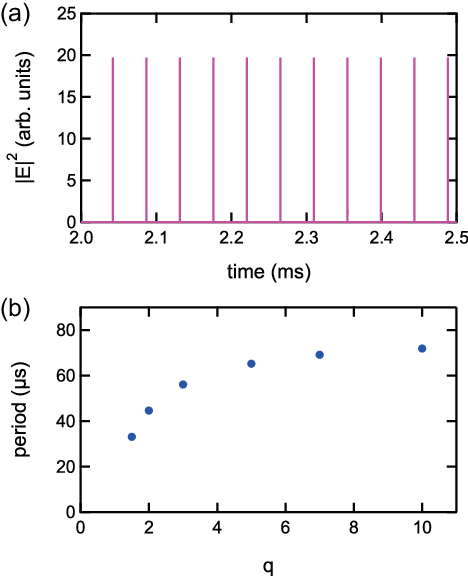}
       \caption{(a) Result of a numerical simulation 
       with $\kappa$ modulation.
       The magenta line is the square of the absolute value 
       of the electric field.
       The parameters are 
       $(P_{13},A_{21},A_{31},A_{32},\gamma_{32})=(100,\underline{2.5\times10^{6}},110,3,10^{8})$ Hz and 
       $(q,E_{\mathrm{th}})=(2,10^{-5} \times (\sqrt{3} \hbar \Omega_{0}/ d_{32}))$.
       (b) Dependence of the period on $q$ 
       in Eq. (\ref{eq:kappa}).
       }
       \label{fig:kappamod}
\end{center}
\end{figure}

When the modulation of $\kappa$ is introduced, 
the periodic SR can occur even under the actual parameter set 
$(P_{13},A_{21},A_{31},A_{32},\gamma_{32})=(\underline{200},\underline{2.5\times10^{6}},\underline{10^{2}},\underline{4},\underline{10^{7}})$ Hz.
In this case, 
coherence and the electric field increase slowly at first.
Then, when $|E_{0}|$ exceeds the threshold $E_{\mathrm{th}}$, 
coherence develops faster than without modulation 
due to the reduced $\kappa$.
In effect, SR pulses are emitted more easily 
for a broader parameter space.
To find the best parameter set, 
we use the following search strategy.
First, we vary the actual parameters 
within the uncertainty range shown in Table \ref{tab:param} 
with the fixed modulation parameters $(q,E_{\mathrm{th}})$.
We then vary 
$q$ while keeping $E_{\mathrm{th}}$ fixed 
to see if we can achieve the experimental result.
The process is repeated many times 
(so the study is of a trial-and-error nature).
We found that the best result gives 
\begin{eqnarray}
 && (T,\Delta T, N_{\mathrm{SR}})=(45 \;\mu \textrm{s}, 14 \;\textrm{ns}, 8.5\times 10^{12}) , 
 \label{eq:kappa mod best result} 
\end{eqnarray}
with the parameter set of  
$(P_{13},A_{21},A_{31},A_{32},\gamma_{32})=(100,\underline{2.5\times10^{6}},110,3,10^{8})$ 
and $(q,E_{\mathrm{th}})= (2, 10^{-5} \times ( \sqrt{3} \hbar \Omega_{0}/ d_{32}))$.
Figure \ref{fig:kappamod} (a) shows the resulting SR pulses 
from the numerical simulation.
Eq. (\ref{eq:kappa mod best result}) should be compared with 
Eq. (\ref{eq:main experimental result}), the observed values: 
our simulation results deviate from the actual ones 
by a factor of $\sim 4$. 
We note that the above simulations are rather insensitive to 
$E_{\textrm{th}}$ as long as 
$E_{\mathrm{th}} < 10^{-4} \times ( \sqrt{3} \hbar \Omega_{0} / d_{32} )$.
We also note that if we allow $P_{13}$ to vary 
outside the uncertainty range, 
we can obtain a much more satisfactory result; 
$(T,\Delta T, N_{\mathrm{SR}})=(137 \;\mu \textrm{s}, 16 \;\textrm{ns}, 7.7 \times 10^{12})$, 
which is realized with the parameter set 
$(P_{13},A_{21},A_{31},A_{32},\gamma_{32})=(30,\underline{2.5\times10^{6}},110,3,10^{8})$ Hz and $(q,E_{\mathrm{th}})= (2, 10^{-5} \times (\sqrt{3} \hbar \Omega_{0}/ d_{32}))$.
Although $P_{13}$ lies outside the uncertainty range, 
it may reasonably take a lower value for the following reason.
After excitation to the $^{4}$I$_{9/2}$ state, 
it is assumed that nonradiative decay leads to population accumulation 
in the lowest Stark level of the $^{4}$I$_{13/2}$ state ($\ket{3}$).
However, if alternative relaxation pathways 
that depopulate $\ket{1}$ without passing through $\ket{3}$ exist 
and their branching ratios are high, 
the effective pumping rate $P_{13}$ would decrease.
Such processes, 
involving energy exchange with the surrounding host material, 
are plausible given the complexity of the solid-state crystals.
\\ 

Figure \ref{fig:kappamod} (b) shows 
the dependence of the period on $q$.
The longer period can be seen for the higher $q$.
The actual value of $q$ is not clear 
and should be determined by experiments.
The quantity $q$ can be quite high in our model in Appendix \ref{app:mechanism}.
The parameter we have chosen ($q=2$) is within a reasonable range, 
since it only doubles the effective length.
Whether or not 
the $\kappa$ modulation is realized in our actual experiment 
also remains unanswered in this study.
We may be able to prove/disprove the proposed mechanism 
by an experiment, for example, with a crystal 
having non-parallel exit/entry surfaces 
or surfaces with an appropriate anti-reflection coating. 
Such an experiment is also left as a future task.

%--------------Sec 7--------------------%
\section{Summary}
\label{sec: Summary}
%--------------Sec 7--------------------%

In this paper, we have performed theoretical and simulation studies 
to investigate the periodic superradiance 
observed in an Er:YSO crystal \cite{Hara-pSR}. 
Our particular motivation is 
to understand the phenomena more quantitatively.  
In particular, we want to reproduce by simulation 
the characteristic SR quantities such as 
the period $T$, the pulse duration $\Delta T$, 
and the SR photon number $N_{\mathrm{SR}}$.  
See Eq. (\ref{eq:main experimental result}) 
for the experimental values.  
For this purpose, 
we first constructed a model based on the Maxwell-Bloch equations. 
This model, X2MB, has a reduced number of energy levels, \textit{i.e.} 
a pair of the SR states and a reservoir, 
and is solved by numerical simulation in this study.
It is found that the model successfully generates periodic SR pulses 
under continuous laser excitation.

The original X2MB model has many parameters, 
and it exhibits periodic SRs for a certain limited region
of the parameter space.
It is not obvious that 
a nonlinear equation system such as the X2MB model 
can exhibit periodic solutions, such as periodic SR pulses, 
under static input parameters.
It is found possible to classify the parameter space 
into two distinct regions, the periodic and non-periodic regions, 
for a given set of parameters.   
The classification is made by analyzing the eigenvalues 
of the linearized X2MB equations around the equilibrium points, 
and is confirmed to be valid by the X2MB numerical simulations.
This analysis reveals that periodic SR can occur 
intrinsically from nonlinear dynamics, 
even with static input parameters.

Having established the analysis tool mentioned above, 
we investigated whether or not 
periodic SR can be realized under the actual experimental conditions. 
We found that 
our experimental parameter set is outside the periodic SR region, 
and that this conclusion holds 
even when we allow the parameters to vary 
over a wide range of uncertainties.
We conclude that some other mechanism(s) is required 
to reproduce the observed periodic SR. 

As an example of such mechanisms, we presented a modulation of 
the field decay constant $\kappa$, \textit{i.e.} 
the electric field emission rate to the outside of the crystal, 
as a function of the electric field strength. 
It is found that the X2MB model with this modification exhibits periodic SR even for the actual parameter set. 
With some parameter adjustments within the uncertainties, 
the best simulation results of $(T, \Delta T, N_{\mathrm{SR}})$ 
turn out to be off by a factor $4$ from the experimental values.  
See Eq.(\ref{eq:kappa mod best result}).
Furthermore, 
the orders of magnitude of the experimental values 
can be reproduced by assuming input parameters 
outside the uncertainty range, 
which can be justified by reasonable physical considerations.
We conclude that the X2MB model with this modification 
gives much more satisfactory results. 
Whether or not the $\kappa$ modulation is realized 
in our actual experiment remains unanswered in this study, 
and its experimental verification/refutation is left as a future task. 

We also derived a simpler model from X2MB. 
This model, T2B, 
is actually a set of nonlinear differential equations with only 
two variables, the coherence and the population differences 
between the two SR states.
A nice feature of the model is that it has an integral: 
indeed, the SR characteristic quantities 
$(T, \Delta T, N_{\mathrm{SR}})$ 
can be expressed analytically with an integral. 
Obtaining analytical solutions allows 
the effects and sensitivities of individual parameters 
to be evaluated without the need for full-scale numerical simulations.
In fact, it is found that the prediction by the analytical expressions 
and the numerical results by X2MB simulations are in good agreement.  
It is clear that 
this model contains essential features of the periodic SR : nonlinearity and coherence.

The periodic SR is an interesting phenomenon
because it seems to be in sharp contrast to 
the stochastic nature of SR. 
The study presented in this paper contributes to deeper understanding 
of the new properties of the SR, 
in particular, 
it reveals the mathematical structures behind this phenomenon.
The study is also important 
for the application of the phenomenon to other systems. 
If the level system is found in real atomic or ionic systems 
that satisfies the necessary condition for the periodic SR, 
it would be possible to predict and prepare a system 
that emits periodic optical pulses 
in response to a CW input without any complex mechanisms 
such as modulation of the field decay rate.
This system has a potential as a light source, 
where coherent optical pulses are periodically generated 
in a simple system without Q-switching.
Periodic SR may be related in a broad sense 
to the dissipative structure theory 
by Prigogine \cite{Prigogine-1978}.
It demonstrates the phenomenon of self-organization 
in a nonequilibrium open system due to energy dissipation.
In our experiment an excitation laser works as an energy input 
driving the system into a far-from-equilibrium state, 
and deexcitation processes are considered as a kind of dissipation.

We believe that the above studies explore 
intriguing aspects 
of the Maxwell-Bloch equations and shed new light on SR physics.

\begin{acknowledgments}
We thank professor K. An for helpful discussions 
and warm encouragement.
This work was supported by JSPS KAKENHI 
(Grants No. JP19H00686, No. JP20H00161, No. JP21H01112, No. JP25K00940, No. JP25K01027, and No. JP25K07337) 
and the Korea Research Foundation (Grant No. 2020R1A2C3009299).
\end{acknowledgments}

\appendix

\section{Derivation of equations}
\label{app:derivation}

In the main text of this paper, 
we introduced the T2B model, 
Eqs. (\ref{eq:dXdtau}) - (\ref{eq:dYdtau}).
Here we derive these equations 
from the X2MB equations, 
Eqs. (\ref{eq:drho11dt}) - (\ref{eq:Maxwell}).

First, to make the equations dimensionless, we introduce 
\begin{equation}
\label{eq:epsilonsr_ttilde}
\epsilon_{\mathrm{SR}} \equiv \frac{\Omega_{s}}{\Omega_{0}}, 
\qquad \tilde{t} \equiv t \Omega_{0}. \\
\end{equation}
As mentioned in the main text, 
we set $\Omega_{s}$ and $\rho_{32}$ as real numbers 
without loss of generality.
Thus $\epsilon_{\mathrm{SR}}$ is a real number.
The dimensionless Maxwell-Bloch equations are given by 
\begin{align}
\label{eq:drho11dttilde}
&\frac{d \rho_{11}}{d \tilde{t}} = - p_{13} \rho_{11} + a_{31} \rho_{33} + a_{21} \rho_{22} , \\
\label{eq:drho22dttilde}
&\frac{d \rho_{22}}{d \tilde{t}} = \epsilon_{\mathrm{SR}} \rho_{32} + a_{32} \rho_{33} - a_{21} \rho_{22}, \\
\label{eq:drho33dttilde}
&\frac{d \rho_{33}}{d \tilde{t}} = - \epsilon_{\mathrm{SR}} \rho_{32} + p_{13} \rho_{11} - (a_{31} + a_{32}) \rho_{33}, \\
\label{eq:drho32dttilde}
&\frac{d \rho_{32}}{d \tilde{t}} = \frac{\epsilon_{\mathrm{SR}}}{2} ( \rho_{33} - \rho_{22} ) - \tilde{\gamma}_{32} \rho_{32}, \\
\label{eq:depsilonSRdttilde}
&\frac{\partial \epsilon_{\mathrm{SR}}}{\partial \tilde{t}} = - \tilde{\kappa} \epsilon_{\mathrm{SR}} + \rho_{32} + \tilde{R}_{\mathrm{sp}} \rho_{33}, 
\end{align}
where $p_{13}=P_{13} / \Omega_{0}$, 
$a_{ij}=A_{ij} / \Omega_{0}$, 
$\tilde{\gamma}_{32} = \gamma_{32} / \Omega_{0}$, 
and $\tilde{\kappa} = \kappa / \Omega_{0}$ 
and $\tilde{R}_{\mathrm{sp}} = R_{\mathrm{sp}} / \Omega_{0}$.

Introducing the new variables 
\begin{align}
\label{eq:u_def01}
&u_{0} \equiv \epsilon_{\mathrm{SR}}, 
\qquad u_{1} \equiv \rho_{32}, \\
\label{eq:u_def23}
&u_{2} \equiv \frac{\rho_{33} - \rho_{22}}{2}, 
\qquad u_{3} \equiv \frac{\rho_{33} + \rho_{22}}{2}, 
\end{align}
and using the relation $\rho_{11} + \rho_{22} + \rho_{33} = 1$, 
Eqs. (\ref{eq:drho11dttilde}) - (\ref{eq:depsilonSRdttilde}) 
are reduced as follows.
\begin{align}
\label{eq:du0dttilde}
&\frac{d u_{0}}{d \tilde{t}} = -b_{00} u_{0} + u_{1}, \\
\label{eq:du1dttilde}
&\frac{d u_{1}}{d \tilde{t}} = u_{0} u_{2} - b_{11} u_{1} , \\
\label{eq:du2dttilde}
&\frac{d u_{2}}{d \tilde{t}} = - u_{0} u_{1} + b_{20} + b_{22} u_{2} + b_{23} u_{3}, \\
\label{eq:du3dttilde}
&\frac{d u_{3}}{d \tilde{t}} = b_{30} + b_{32} u_{2} + b_{33} u_{3} ,
\end{align}
where 
\begin{align}
\label{eq:b00b11}
& b_{00} \equiv \tilde{\kappa}, 
\qquad b_{11} \equiv \tilde{\gamma}_{32} , \\
\label{eq:b20b22}
& b_{20} \equiv \frac{p_{13}}{2} , 
\quad b_{22} \equiv \frac{- a_{31} - a_{21} - 2 a_{32}}{2}, \\
\label{eq:b23b30}
& b_{23} \equiv \frac{- 2 p_{13} - a_{31} + a_{21} - 2 a_{32}}{2}, 
\quad b_{30} \equiv \frac{p_{13}}{2} , \\
\label{eq:b32b33}
& b_{32} \equiv \frac{- a_{31} + a_{21}}{2}, 
\quad b_{33} \equiv \frac{- 2 p_{13} - a_{31} - a_{21}}{2}.
\end{align}
The coherence trigger term 
in Eq. (\ref{eq:depsilonSRdttilde}) is ignored here.

When the time derivatives 
in Eqs. (\ref{eq:du0dttilde}) - (\ref{eq:du3dttilde}) 
are set to zero, 
$u_{i}$ at the equilibrium point are given by 
\begin{align}
\label{eq:u0e}
& u_{0}^{e} = b_{00}^{-1} u_{1}^{e} , \\
\label{eq:u1e}
& u_{1}^{e} = \pm \sqrt{b_{00} \Big( b_{20} + b_{00} b_{11} b_{22} - \frac{b_{23}}{b_{33}} ( b_{30} + b_{00} b_{11} b_{32} ) \Big)} , \\
\label{eq:u2e}
& u_{2}^{e} = b_{00} b_{11}, \\
\label{eq:u3e}
& u_{3}^{e} = - \frac{b_{30} + b_{00} b_{11} b_{32}}{b_{33}}.
\end{align}
There is also the following trivial solution.
\begin{align}
\label{eq:u01etriv}
& u_{0}^{e,\mathrm{triv}} = u_{1}^{e,\mathrm{triv}} =0 , \\
\label{eq:u2etriv}
& u_{2}^{e,\mathrm{triv}} = - \frac{b_{33} b_{20} - b_{23} b_{30}}{b_{22} b_{33} - b_{23} b_{32}}, \\
\label{eq:u3etriv}
& u_{3}^{e,\mathrm{triv}} = \frac{b_{32} b_{20} - b_{22} b_{30}}{b_{22} b_{33} - b_{23} b_{32}}. 
\end{align}

So far we have rewritten the X2MB equations 
into a dimensionless form.
Below we derive the T2B equations using several approximations.
In Eq. (\ref{eq:du0dttilde}), the time derivative of $u_{0}$ 
can be ignored because $b_{00}$, 
which corresponds to the field decay rate $\kappa$, 
is very high and dominates the time evolution of the system.
Using this approximation, $u_{0}=b_{00}^{-1} u_{1}$, 
the electric field is proportional to the coherence.
By introducing the following new variables 
\begin{equation}
X \equiv \frac{u_{1}}{u_{1}^{e}}, 
\qquad Y \equiv \frac{u_{2} - u_{2}^{e}}{u_{1}^{e}}, 
\qquad \tau \equiv \frac{u_{1}^{e}}{b_{00}} \tilde{t} ,
\end{equation}
Eq. (\ref{eq:du1dttilde}) is reduced to 
\begin{equation}
\frac{d X}{d \tau} = X Y .
\end{equation}

Equation (\ref{eq:dYdtau}) is derived as follows.
We first transform Eqs. (\ref{eq:du2dttilde}) 
and (\ref{eq:du3dttilde}) into the following form: 
\begin{align}
\label{eq:du2dttildeeq}
\frac{d u_{2}}{d \tilde{t}} = & c_{21} \bigg\{ \bigg( \frac{u_{1}}{u_{1}^{e}} \bigg)^{2} - 1\bigg\} + c_{22} \bigg( \frac{u_{2}}{u_{2}^{e}} - 1 \bigg) + c_{23} \bigg( \frac{u_{3}}{u_{3}^{e}} -1 \bigg), \\
\label{eq:du3dttildeeq}
\frac{d u_{3}}{d \tilde{t}} = & c_{32} \bigg( \frac{u_{2}}{u_{2}^{e}} - 1 \bigg) + c_{33}  \bigg( \frac{u_{3}}{u_{3}^{e}} -1 \bigg) , 
\end{align}
where
\begin{align}
\label{eq:dudt c21}
c_{21} = & -\frac{(u_{1}^{e})^{2}}{b_{00}} & 
\to \quad & a_{21}\frac{1}{1+\eta} & 
\sim & \quad \mathcal{O}(10^{-8}),  \\  
\label{eq:dudt c22}
c_{22} = & b_{22} u_{2}^{e} & 
\to \quad & a_{21} \tilde{\gamma}_{32} \tilde{\kappa} & 
\sim & \quad \mathcal{O}(10^{-12}), \\
\label{eq:dudt c23}
c_{23} = & b_{23} u_{3}^{e} & 
\to \quad & p_{13} \frac{1-\eta}{1+\eta} & 
\sim &  \quad \mathcal{O}(10^{-7}), \\
\label{eq:dudt c32}
c_{32} = & b_{32} u_{2}^{e} & 
\to \quad & a_{21} \tilde{\gamma}_{32} \tilde{\kappa} & 
\sim & \quad \mathcal{O}(10^{-12}), \\ 
\label{eq:dudt c33}
c_{33} = & b_{33} u_{3}^{e} & 
\to \quad & p_{13} & 
\sim & \quad \mathcal{O}(10^{-7}), 
\end{align}
with $\eta\equiv a_{21}/2p_{13}$.
On the right of the arrows above, 
major parameter dependence is shown 
considering the hierarchy of Eq. (\ref{eq:hierarchy}). 
Here we note $\tilde{\kappa} \simeq 1$ 
and the rest ($\tilde{\gamma}_{32}, p_{13}, a_{21}$) 
are $10^{-3}$ or less, 
indicating that $c_{22}$ and $c_{32}$ 
are much smaller than $c_{21}, c_{23}, c_{33}$.
Similarly, the equilibrium population $u_{3}^e$ 
can be expressed as 
\begin{align}
\label{eq:2u3e app}
2u_{3}^e=\rho_{22}^e+\rho_{33}^e \quad \to \quad \frac{1}{1+\eta} 
\quad \sim \quad 0.95. %\mathcal{O}(1).
\end{align}
The values inside the symbol of $\mathcal{O}$ above 
represent those obtained by the X2MB simulation 
at the purple diamond point in Fig. \ref{fig:stability}. 
Now if $u_3$ stays near $u_{3}^e$ or $(u_3/u_3^e)-1 \ll 1$, 
then the right-hand side of Eq.(\ref{eq:du3dttildeeq}) 
becomes negligible because $c_{32} \ll 1$, or 
\begin{equation}
\label{eq:du3dtau}
\frac{d u_{3}}{d \tau} \sim 0, 
\end{equation}
which in turn guarantees $u_3$ to be constant. 
In addition, as $2u_{3}^e\sim 1$ for $\eta<1$, 
the population of $\rho_{11}$ is almost empty.
In fact, we found by the X2MB simulation 
$(u_{3}/u_{3}^{e})-1 \sim \mathcal{O} (10^{-6})$, 
$(u_{2}/u_{2}^{e})-1 \sim \mathcal{O} (1)$, 
and $(u_{1}/u_{1}^{e})^{2}-1 \sim \mathcal{O} (10^{2})$ 
at the purple diamond point, 
and confirmed similar results are obtained 
at seven different points in the periodic SR region 
of Fig. \ref{fig:stability}.
With the above approximation, 
the second and third terms in Eq.(\ref{eq:du2dttildeeq}) 
are negligible compared to the first term. 
After appropriate variable transformations, we obtain 
\begin{equation}
\frac{dY}{d \tau} = - (X^{2} - 1). \\
\end{equation}
The T2B model, 
Eqs. (\ref{eq:dXdtau}) and (\ref{eq:dYdtau}), 
is derived by the above procedure.

\section{Analysis using rate equations}
\label{app:rateeq}

\begin{figure}[t]
\begin{center}
      \includegraphics[width=7.5cm]{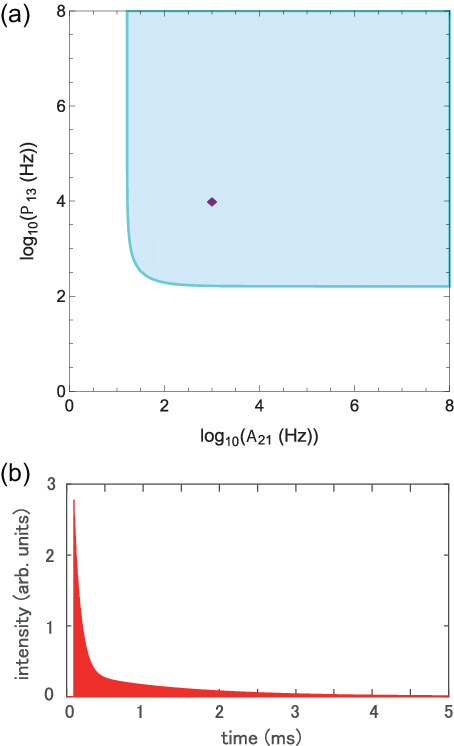}
       \caption{(a) Region of $(u'_{1})^{e}>0$ 
       in the $(A_{21},P_{13})$ plane 
       with $(A_{31},A_{32})$ fixed at $(\underline{10^{2}}, \underline{4})$ Hz.
       The purple diamond represents the parameter 
       used in the simulation of (b).
       (b) Numerical simulation result for the parameter set 
       $(P_{13},A_{21},A_{31},A_{32})=(10^{4},10^{3},\underline{10^{2}},\underline{4})$ Hz.
       The red line is 
       the intensity of the generated optical pulses.
       }
       \label{fig:rateeq}
\end{center}
\end{figure}

Here we show that the rate equations 
used for relaxation oscillation in laser systems 
only give a damped oscillation solution.
For the Maxwell-Bloch Eqs. (\ref{eq:drho11dt})-(\ref{eq:Maxwell}), 
neglecting coherence 
and replacing the Rabi oscillation terms 
with stimulated emission/absorption terms, 
we obtain the following rate equations.
\begin{align}
\label{eq:drho11dt_rateeq}
\frac{d \rho_{11}}{d t} & = -P_{13} \rho_{11} + A_{31} \rho_{33} + A_{21} \rho_{22} , \\
\label{eq:drho22dt_rateeq}
\frac{d \rho_{22}}{d t} & = (\rho_{33} - \rho_{22}) \int B_{32} \rho (\omega) g (\omega) d \omega + A_{32} \rho_{33} - A_{21} \rho_{22}, \\
\label{eq:drho33dt_rateeq}
\frac{d \rho_{33}}{d t} & = - (\rho_{33} - \rho_{22}) \int B_{32} \rho (\omega) g (\omega) d \omega \notag \\
& + P_{13} \rho_{11} - (A_{31} + A_{32}) \rho_{33}, \\
\label{eq:intensity}
\frac{\partial I}{\partial t} & = - 2 \kappa I + \frac{\hbar \omega_{32} N_{0} c}{n_{0}} ( \rho_{33} - \rho_{22} ) \int B_{32} \rho (\omega) g (\omega) d \omega \notag \\
& + R_{\mathrm{sp}}'\frac{\hbar \omega_{32} N_{0} c}{n_{0}} A_{32} \rho_{33} .
\end{align}
Here $I$ is the intensity of the generated optical pulse.
$B_{32}=\frac{\pi^{2} c^{3}}{n_{0}^{3} \hbar \omega_{32}^{3}} A_{32}$ is the Einstein $B$ coefficient.
$\rho(\omega) = (n_{0}/c) I f(\omega)$ 
is the energy density per unit frequency, 
where $f(\omega)$ is the spectral function.
$g (\omega)$ is the ionic lineshape function.
For simplicity, $f(\omega)$ and $g(\omega)$ are approximated 
by rectangular functions 
with widths of $2 \pi \times 0.1$ GHz and $2 \pi \times 1$ GHz, 
respectively.
These widths are set based on the measurement.
The last term of Eq. (\ref{eq:intensity}) 
is a contribution to the generated optical pulse 
by spontaneous emissions.
The rate $R_\mathrm{sp}'$ is extremely small.

Using the relation $\rho_{11} + \rho_{22} + \rho_{33} = 1$ 
and the appropriate variable transformations, 
Eqs. (\ref{eq:drho11dt_rateeq}) - (\ref{eq:intensity}) 
are reduced as follows.
\begin{align}
\label{eq:du1dttilde_rateeq}
&\frac{d u'_{1}}{d \tilde{t}} = - b_{11}' u'_{1} + b_{12}' u'_{1} u'_{2} , \\
\label{eq:du2dttilde_rateeq}
&\frac{d u'_{2}}{d \tilde{t}} = - b_{21}' u'_{1} u'_{2} + b_{20}' + b_{22}' u'_{2} + b_{23}' u'_{3}, \\
\label{eq:du3dttilde_rateeq}
&\frac{d u'_{3}}{d \tilde{t}} = b_{30}' + b_{32}' u'_{2} + b_{33}' u'_{3} ,
\end{align}
where 
\begin{align}
\label{eq:t_u1_def}
&\tilde{t} \equiv t \Omega_{0}, \quad u'_{1} \equiv I, \\
\label{eq:u2_u3_def}
&u'_{2} \equiv \frac{\rho_{33} - \rho_{22}}{2}, 
\quad u'_{3} \equiv \frac{\rho_{33} + \rho_{22}}{2}. 
\end{align}
The last term of Eq. (\ref{eq:intensity}) is ignored here.
If the obtained 
Eqs. (\ref{eq:du1dttilde_rateeq})-(\ref{eq:du3dttilde_rateeq}) 
are expanded around the nontrivial equilibrium point 
$((u'_{1})^e, (u'_{2})^{e}, (u'_{3})^{e})$, 
the following linearized differential equations are derived.
\begin{equation}
\label{eq:linearized_rateeq}
\frac{d \vec{y'}}{d \tilde{t}} = J' \vec{y'}, 
 J'= \begin{pmatrix}
    -b_{11}' + b_{12}' (u'_{2})^{e} & b_{12}' (u'_{1})^{e} & 0 \\
    - b_{21}' (u'_{2})^{e} & b_{22}' - b_{21}' (u'_{1})^{e} & b_{23}' \\
    0 & b_{32}' & b_{33}' \\
    \end{pmatrix} , 
\end{equation}
where $J'$ is a Jacobi matrix 
and $y'_{i} \equiv u'_{i} - (u'_{i})^{e} \quad (i=1,2,3)$ 
is a deviation from the equilibrium point.

Figure \ref{fig:rateeq} (a) shows 
the region of $(u'_{1})^{e} > 0$.
This means that the intensity at the equilibrium point is positive.
In this region it is revealed that one of the eigenvalues of $J'$ 
is a negative real number 
and the other two are imaginary numbers with negative real parts.
The system is expected to reach equilibrium 
and no longer generate a sustained pulse train.
The above mathematical discussion is also presented 
for relaxation oscillation in laser systems \cite{Siegman-1986}.
Figure \ref{fig:rateeq} (b) shows the intensity of the optical pulses generated.
We choose the same parameter set as in the simulation 
in Fig. \ref{fig:simulation} (a).
That is, 
$(P_{13},A_{21},A_{31},A_{32})=(10^{4},10^{3},\underline{10^{2}},\underline{4})$ Hz.
It is represented by the purple diamond 
in Fig. \ref{fig:rateeq} (a).
Initially, many optical pulses are generated, 
but as expected, the system reaches equilibrium 
and no longer generates intense pulses.
In the region of $(u'_{1})^{e}<0$, the equilibrium point is 
outside the physically meaningful region.
It is natural to assume that there are no solutions 
orbiting around the equilibrium point.
We also performed numerical simulations 
within the region of $(u'_{1})^{e}<0$ 
and confirmed that the sustained pulses are not reproduced.

As shown above, 
the Maxwell-Bloch equations with coherence 
and the rate equations without coherence yield 
the different results.
These results indicate that 
coherence plays an important role in our experimental system 
and the phenomenon we are considering is different 
from the relaxation oscillation in laser systems.

\section{Comparison of the analytical solution of the T2B model and numerical simulation of the X2MB model}
\label{sec:comparison}

\begin{figure*}[t!]
\begin{center}
      \includegraphics[width=17cm]{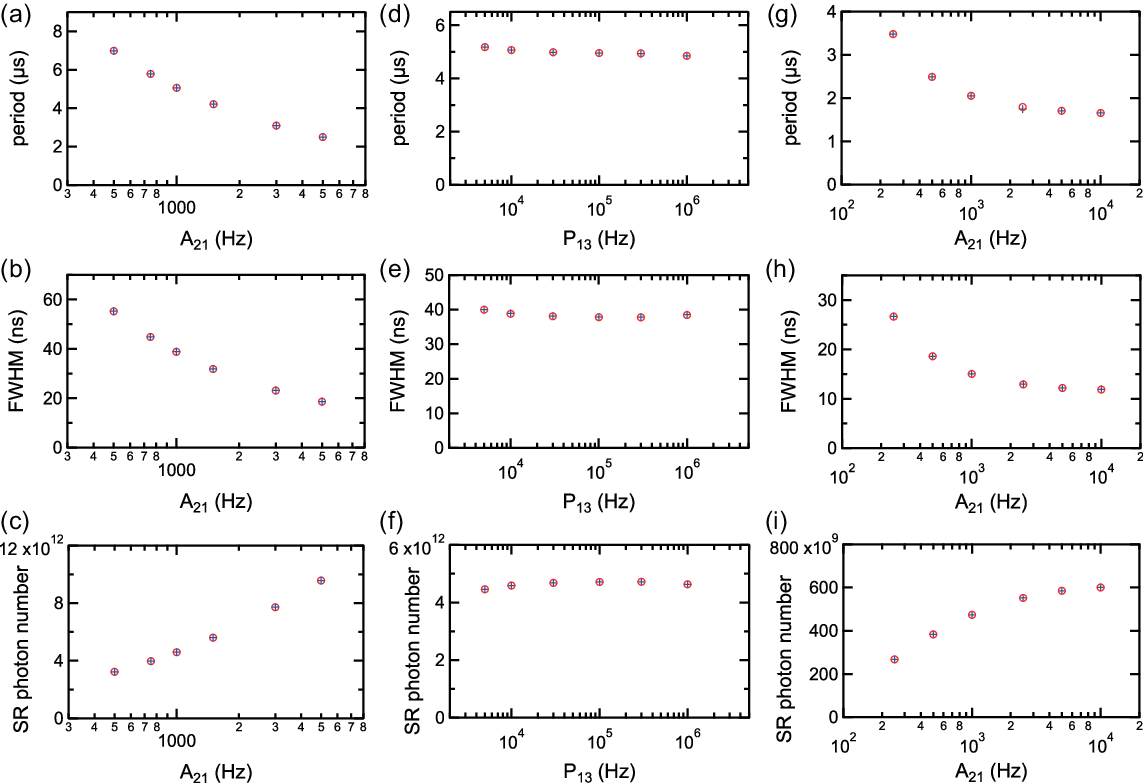}
       \caption{Comparison of the analytical solution 
       of the T2B model and the numerical simulation 
       of the X2MB model.
       The left column shows 
       (a) period, (b) FWHM pulse duration, (c) SR photon number 
       for the different decay rates $A_{21}$ under 
       $(P_{13}, A_{31}, A_{32}, \gamma_{32})=(10^{4},\underline{10^{2}},\underline{4},\underline{10^{7}})$ Hz.
       The middle column shows 
       (d) period, (e) FWHM pulse duration, (f) SR photon number 
       for the pumping rates $P_{13}$ under 
       $(A_{21}, A_{31}, A_{32}, \gamma_{32})=(10^{3},\underline{10^{2}},\underline{4},\underline{10^{7}})$ Hz.
       The right column shows 
       (g) period, (h) FWHM pulse duration, (i) SR photon number 
       for the decay rate $A_{21}$ under 
       $(P_{13}, A_{31}, A_{32}, \gamma_{32})=(\underline{200},\underline{10^{2}},100,\underline{10^{7}})$ Hz.
       The blue crosses and red circles represent the results of 
       the analytical solution 
       and the numerical simulation, respectively.}
       \label{fig:comparison}
\end{center}
\end{figure*}

To confirm the validity of the T2B model 
Eqs. (\ref{eq:dXdtau}) - (\ref{eq:dYdtau}) 
derived with several approximations, 
we compare the analytical solution of the T2B model 
and the numerical simulation result of the X2MB model 
for the same parameter set.
We vary $A_{21}$ or $P_{13}$ in the periodic SR region 
while keeping the other parameters fixed.
In the numerical simulation, we take the values 
after the period and pulse shape have become constant.
As a point of caution, 
the analytical solution requires the prediction of 
the integration constant $C$.
We use the maximum value of coherence $\rho_{\mathrm{max}}$ 
in the numerical simulation to find $C$.
Therefore, one numerical simulation 
is necessary to obtain one analytical solution.
With $C$ above, 
the period, FWHM pulse duration, and SR photon number 
are obtained.

The left column of Fig. \ref{fig:comparison} shows 
(a) period, (b) FWHM pulse duration, and (c) SR photon number 
for the different decay rates $A_{21}$ 
under 
$(P_{13}, A_{31}, A_{32}, \gamma_{32})=(10^{4},\underline{10^{2}},\underline{4},\underline{10^{7}})$ Hz.
This corresponds to the variation on the horizontal line 
$P_{13}=10^{4}$ Hz in the green region of Fig. \ref{fig:stability}.
The middle column of Fig. \ref{fig:comparison} 
shows the results for the different pumping rates $P_{13}$ under 
$(A_{21}, A_{31}, A_{32}, \gamma_{32})=(10^{3},\underline{10^{2}},\underline{4},\underline{10^{7}})$ Hz.
This corresponds to the variation on the vertical line 
$A_{21}=10^{3}$ Hz in the green region of Fig. \ref{fig:stability}.
The right column of Fig. \ref{fig:comparison} shows the results 
for the decay rate $A_{21}$ under 
$(P_{13}, A_{31}, A_{32}, \gamma_{32})=(\underline{200},\underline{10^{2}},100,\underline{10^{7}})$ Hz.
This corresponds to the variation on the horizontal line 
$P_{13}=200$ Hz in the red region of Fig. \ref{fig:stability}.
The blue crosses and red circles represent 
the results of the analytical solution 
of the T2B model 
and the numerical simulation of the X2MB model, 
respectively.
They are in good agreement.
This result shows the validity of several approximations 
used in the T2B model.
The FWHM pulse duration, $\mathcal{O} (10)$ ns, 
and the SR photon number, $\mathcal{O} (10^{12\pm 1})$, 
are comparable with those in the experimental data.
The period, $\mathcal{O} (1)$ $\mu$s, 
is two orders of magnitude shorter than that in the experiment.
As $A_{21}$ increases, the period becomes shorter. 
This is because as $A_{21}$ increases, 
the population inversion reaches the threshold of SR earlier.
On the other hand, as $A_{21}$ increases, 
the FWHM pulse duration becomes shorter 
and the number of SR photons increases.
The FWHM pulse duration is inversely proportional to 
the SR photon number, 
which is one of the characteristics of SR.
The fact that the results change little 
when $P_{13}$ is varied suggests 
that the rate-limiting factor of change in the system 
is due to other parameters.

\section{Possible mechanism of field decay rate modulation}
\label{app:mechanism}

In Sec. \ref{sec:other}, 
we perform the numerical simulation 
assuming that the field decay rate $\kappa$ varies 
as a function of the electric field.
A possible mechanism for dynamically varying $\kappa$ 
is presented here.
In summary, the electric field due to photons 
emitted from the Er$^{3+}$ ions creates a standing wave in the crystal, 
resulting in a modulation of the refractive index.
The refractive index modulation increases the effective reflectance $R$.
In our model, $R$ can be quite high.
Taking into account the propagation loss of the light in the crystal, 
$\kappa$ can be approximated to a binary function 
as in Eq. (\ref{eq:kappa}).
The details are described below.

\subsection{Principle}
\label{subsec:principle}

\subsubsection{Formation of refractive index modulation}
\label{parag:nmod}

The population of $\ket{3}$ in Fig. \ref{fig:threelevel} 
is gradually increased by the pumping $P_{13}$.
The excited Er$^{3+}$ ions deexcite to $\ket{2}$ 
with photons, which generate an electric field.
The reflection of the electric field at the two crystal surfaces 
creates a very low finesse cavity. 
The surface reflection, estimated from the refractive index, is 8 \%.
A small standing wave is formed 
and the electric field intensity is periodically modulated.
At the anti-nodes the population of $\ket{3}$ decreases 
faster than at the nodes 
due to the higher stimulated emission rate.
The electric field at the anti-nodes is further increased.
The refractive index of the crystal $n$ changes 
in response to an electric field due to the optical Kerr effect.
\begin{equation}
n = n_{0} + 2 \tilde{n}_{2} |E|^{2} ,
\end{equation}
where $E$ is the electric field, 
$n_{0}$ is the refractive index in the low-intensity limit, 
and $\tilde{n}_{2}$ is the optical Kerr coefficient.
Therefore, the refractive index modulation is generated.

\subsubsection{Decrease in field decay rate due to increased finesse}
\label{parag:finesse}

\begin{figure}[t]
\begin{center}
       \includegraphics[width=8.5cm]{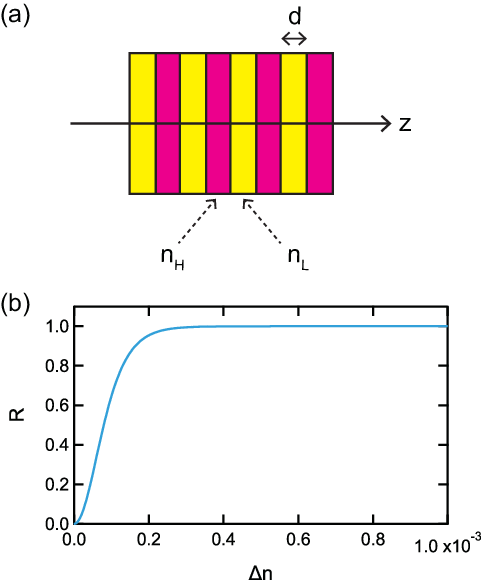}
       \caption{(a) Multilayer structure whose refractive index has 
       two values of $n_{\mathrm{H}}$ and $n_{\mathrm{L}}$, 
       changing periodically with a period of $d$ along the $z$-axis.
       (b) Dependence of the reflectance $R$ 
       on the refractive index variation $\Delta n$.
      }
       \label{fig:multilayer}
\end{center}
\end{figure}

For simplicity, we assume a medium 
whose refractive index has two values 
of $n_\mathrm{H}$ and $n_\mathrm{L}$, 
which change periodically with a period of $d$ along the $z$-axis, 
as shown in Fig. \ref{fig:multilayer} (a).
We define the mean and difference index as
\begin{eqnarray}
 && \bar{n}=\frac{n_\mathrm{H}+n_\mathrm{L}}{2}, \hspace{2cm}
 \Delta n=\frac{n_\mathrm{H}-n_\mathrm{L}}{2} . 
 \label{eq:nbarDeltan} 
\end{eqnarray}
The wavelength of the electromagnetic wave 
propagating in the medium 
is $\lambda_{0} =4 \bar{n} d$ when measured in vacuum.
The transfer matrix \cite{fowles-1989} for the region 
$j\;(j=\mathrm{H},\mathrm{L})$ is given by 
\begin{align}
\label{eq:transfer_matrix}
&\mathcal{M}_{j} (\Delta z) \equiv \begin{pmatrix}
    \cos (k_{j} \Delta z) & \frac{i}{n_{j}} \sin (k_{j} \Delta z) \\
    i n_{j} \sin (k_{j} \Delta z) & \cos (k_{j} \Delta z) \\
    \end{pmatrix} , \\
\label{eq:trasfer_EcB}
&\begin{pmatrix}
   E (z + \Delta z) \\
   c B (z + \Delta z) \\
    \end{pmatrix}
  = \mathcal{M}_{j} (\Delta z) 
  \begin{pmatrix}
   E (\Delta z) \\
   c B (\Delta z) 
    \end{pmatrix}, 
\end{align}
where 
$k_j=n_j k_0$ is the wave vector of the photon 
traveling through the crystal.
%and set $\bar{n}k_0 d=\pi/2$.
Now we look at the multilayer structure again.
If we denote the input and output fields by 
$E_{\mathrm{in}}$, $c B_{\mathrm{in}}$ 
and $E_{\mathrm{out}}$, $c B_{\mathrm{out}}$, respectively, 
they are connected by the following relationship.
\begin{align}
\label{eq:trasfer_inout}
&\begin{pmatrix}
   E_{\mathrm{in}} \\
   c B_{\mathrm{in}} \\
    \end{pmatrix}
  = \big( \mathcal{M}_{\mathrm{H}} (d) \mathcal{M}_{\mathrm{L}} (d) \big)^{- N_{p}} 
  \begin{pmatrix}
   E_{\mathrm{out}} \\
   c B_{\mathrm{out}} 
    \end{pmatrix} .
\end{align}
The number of the index pairs $N_{p}$ can be roughly estimated as 
\begin{equation}
\label{eq:Np}
N_{p} = \frac{\bar{n} L}{\lambda_{0} / 2} \sim 10^{4} ,
\end{equation}
for the mean refractive index $\bar{n}=1.8$, 
the target length $L=6$ mm, 
and the wavelength of SR $\lambda_{0}=1.545$ $\mu$m. 
If the incident beam is injected from the left, 
then there are both right-moving (injected) 
and left-moving (reflected) beams at the input boundary, 
but there is no left-moving field at the output boundary.
Denoting the incident field by $E_{1}$, 
the reflected field by $E_{\mathrm{R}}$, 
and the transmitted field by $E_{\mathrm{T}}$, 
we have 
\begin{align}
\label{eq:EBin}
& E_{\mathrm{in}} = E_{1} + E_{\mathrm{R}}, 
\quad c B_{\mathrm{in}} = E_{1} - E_{R},  \\
\label{eq:EBout}
& E_{\mathrm{out}} = E_{\mathrm{T}} , 
\quad c B_{\mathrm{out}} = E_{\mathrm{T}} . 
\end{align}
Denoting the matrix elements of 
$(\mathcal{M}_{\mathrm{H}} \mathcal{M}_{\mathrm{L}})^{-N_{p}}$ 
by $m_{ij}$, 
we obtain the complex reflectivity $r$ as
\begin{eqnarray}
 && r\equiv \frac{E_R}{E_1}=\frac{\displaystyle (m_{11}+ m_{12})-(m_{21}+ m_{22})}
 {\displaystyle \displaystyle (m_{11}+m_{12})+(m_{21}+ m_{22})} .
 \label{eq:r and t amplitude ratio-again} 
\end{eqnarray}
Figure \ref{fig:multilayer} (b) shows $R(=|r|^{2})$ 
as a function of $\Delta n$.
There seems to be a threshold 
around $\Delta n\sim 1\times 10^{-4}$, 
above which $R$ is close to 1.
This is a highly nonlinear effect.
For a cavity with reflectivity $R$, length $l=2L$, 
and energy reduction $e^{-\alpha l}$ in a round trip, 
we can write the electric field decay rate 
$\kappa = \kappa_{0} \times (\alpha l - 2 \ln{R})$, 
where $\kappa_{0}=c/n_{0} L$ is the original field decay rate.
It is natural to assume that $\kappa / \kappa_{0}$ is lower than 1, 
\begin{equation}
\kappa / \kappa_{0} = \min{\Big[1,\alpha l - 2 \ln{R}\Big]}.
\label{eq:q}
\end{equation}
The 1 here is a conceptual number 
determined from the reciprocal of the time 
it takes for a photon to propagate through the target, 
and is an approximate value.
$\kappa / \kappa_{0}$ can be viewed as a function with respect to an electric field 
that varies from 1 to $\sim \alpha l$ (because $R \sim 1$).
This functional form can be approximated 
by a tanh function and further by a step function.
By setting $1/q = \alpha l$, Eq. (\ref{eq:kappa}) is derived.
In fact, it is difficult to determine the loss $\alpha$.
This is equivalent to determining 
the higher value of the step function.
The threshold electric field can be 
determined by the detuning $\delta$, 
in Appendix \ref{subsec:Deltan}.
The field decay rate $\kappa$ decreases by a factor of $q$.
This is the possible mechanism 
by which the field decay rate $\kappa$ varies 
as a function of the electric field.
For the strong electric field, 
the refractive index variation $\Delta n$ increases 
and the field decay rate $\kappa$ decreases significantly.

\subsection{Refractive index variation $\Delta n$}
\label{subsec:Deltan}

We estimate the refractive index variation $\Delta n$ 
for a given electric field $E_{\mathrm{SR}}$.
According to the results of Ref. \cite{Boyd-textbook-2008}, 
in particular Eq.(6.3.28)
\footnote{It assumes a near-resonant, steady-state two-level system.}, 
the electric susceptibility  $\chi_e$ is given by 
\begin{align}
 & \chi_e=\frac{-\alpha_0(0)}{\omega_{32}/c}\frac{\delta T_2-i}{1+(\delta T_2)^2+|E|^2/|E_S^0|^2} ,
 \label{eq:Boyd chi}  \\ 
 & |E_S^0|^2=\frac{\hbar^2}{4|d_{32}|^2T_1T_2}, \\
 &\alpha_0(0)=-\frac{\omega_{32}}{c}\left[N_{0} (\rho_{33}-\rho_{22})|d_{32}|^2\frac{T_2}{\epsilon_0 \hbar}\right],
\end{align}
where $\alpha_0(0)$ is the absorption coefficient at zero detuning, 
$N_{0}$ is the number density of Er$^{3+}$ ions at site 2, 
$\rho_{ij}$ is the density matrix element, 
$d_{32}$ is the dipole moment, 
$\delta$ is the detuning, 
$T_{1}$ is the lifetime of $\ket{3}$, 
and $T_{2}$ is the decoherence time.
The third-order susceptibility $\chi_{e}^{(3)}$ has 
the relation $\tilde{n}_{2} = 3 \chi_{e}^{(3)} / 4 n_{0}$ 
with $\tilde{n}_{2}$. 
Expanding Eq. (\ref{eq:Boyd chi}) with respect to $|E|/|E_{S}^{0}|$, 
the third-order susceptibility $\chi_{e}^{(3)}$ is 
\begin{equation}
  \chi_{e}^{(3)} = \frac{\alpha_{0} (0)}{3 \omega_{32} / c} \frac{\delta T_{2}}{[1+(\delta T_{2})^{2}]^{2}} \frac{1}{|E_{S}^{0}|^{2}} .
 \label{eq:chi3}
\end{equation}
The number density of Er$^{3+}$ ions 
relevant to the index variation $\Delta n$ is 
given by the SR electric field $E_{\mathrm{SR}}$. 
The number density can be replaced by 
\begin{equation}
  N_{0} (\rho_{33} - \rho_{22}) \sim \frac{\epsilon_0 n_{0}^{2} |E_{\mathrm{SR}}|^2}{\hbar \omega_{32}}.
 \label{eq:replace} 
\end{equation}
With this substitution, the refractive index variation $\Delta n$ 
is estimated by 
\begin{align}
 \Delta n & \sim 2\times \frac{3}{4 n_{0}} \big| \Delta \chi_{e}^{(3)} \big| \times \big|E_{\mathrm{SR}} \big|^{2}\\ 
 & \sim \left[\frac{\epsilon_0 n_{0} |E_{\mathrm{SR}}|^2}{2 \hbar \omega_{32}}|d_{32}|^2\frac{T_2}{\epsilon_0 \hbar}\right]
 \frac{\delta T_2}{[1+(\delta T_2)^2]^2} \frac{|E_{\mathrm{SR}}|^{2}}{|E_{S}^{0}|^{2}} .
 \label{eq:Deltan_variation}
\end{align}
To estimate $\Delta n$, we need to know several parameters, 
of which $\delta$ and $E_{\mathrm{SR}}$ are the least well known.
Naively $\delta$ is expected to be smaller than the width of 
the Er$^{3+}$ ion ensemble, \textit{i.e.} $\delta < 10^{9}$ Hz.
Remembering that $\Delta n \sim 10^{-4}$ is the threshold for 
the sharp increase in $R$ or the sharp decrease in $\kappa$, 
we evaluate $E_{\mathrm{SR}}$ that gives $\Delta n = 10^{-4}$.
It turns out that 
\begin{align}
 |E_{\mathrm{SR}}|
\sim 6\times 10^{5} \quad \mbox{V/m} ,
 \label{eq:Esr needed for dn of 10-4} 
\end{align}
for $\delta = 10^{9}$ Hz.
The number of photons giving this electric field is estimated to be 
\begin{align}
 \frac{\epsilon_0 n_{0}^{2} |E_{SR}|^2}{\hbar \omega_{32}} \times \pi w_{0}^{2} \times \frac{c}{n_{0}} \times \Delta T 
\sim 1 \times 10^{13} .
 \label{eq:NSR_Deltan} 
\end{align}
$\Delta T$ is the duration of the SR pulse 
and is here set to $10^{-8}$ s.
On the other hand, the electric field estimated 
from the measured SR pulses is $\sim 2\times 10^{5}$ V/m.
We note that if we assume $\delta = 10^{8}$ Hz, 
then we find $|E_{\mathrm{SR}}| \sim 1 \times 10^{5}$ V/m 
and the corresponding photon number $\sim 3 \times 10^{11}$, 
much smaller than those of Eqs.(\ref{eq:Esr needed for dn of 10-4}) 
and (\ref{eq:NSR_Deltan}).
These considerations suggest 
that the modulation of the field decay rate 
$\kappa$ may occur.

%\bibliography{SRsim}


\end{document}